\documentclass[sn-mathphys,Numbered]{sn-jnl}

\usepackage{graphicx}%
\usepackage[skip=14pt]{caption}
\usepackage{subcaption}
\usepackage{multirow}%
\usepackage{amsmath,amssymb,amsfonts}%
\usepackage{amsthm}%
\usepackage{mathrsfs}%
\usepackage[title]{appendix}%
\usepackage{xcolor}%
\usepackage{textcomp}%
\usepackage{manyfoot}%
\usepackage{booktabs}%
\usepackage{algorithm}%
\usepackage{algorithmicx}%
\usepackage{algpseudocode}%
\usepackage{listings}%
\usepackage{soul}%
\usepackage{lipsum}
\usepackage{url}

\usepackage{afterpage}
\usepackage{geometry}
\usepackage{makecell}

\begin{document}

\title[Article Title]{Classical Interfaces for Controlling Cryogenic Quantum Computing Technologies}


\author*[1]{\fnm{Jack C.} \sur{Brennan}}\email{jack.brennan@glasgow.ac.uk}

\author[1]{\fnm{Jo\~ao} \sur{Barbosa}}

\author[1]{\fnm{Chong} \sur{Li}}

\author[1]{\fnm{Meraj} \sur{Ahmad}}

\author[1]{\fnm{Fiheon} \sur{Imroze}}

\author[1]{\fnm{Calum} \sur{Rose}}

\author[1]{\fnm{Wridhdhisom} \sur{Karar}}

\author[2]{\fnm{Manoj} \sur{Stanley}}

\author[1]{\fnm{Hadi} \sur{Heidari}}

\author[2]{\fnm{Nick M.} \sur{Ridler}}

\author[1]{\fnm{Martin} \sur{Weides}}

\affil[1]{\orgdiv{James Watt School of Engineering}, \orgname{University of Glasgow}, \orgaddress{\street{Advanced Research Centre}, \city{Glasgow}, \postcode{G11 6EW}, \state{Scotland}, \country{United Kingdom}}}

\affil[2]{\orgdiv{Electromagnetic \& Electrochemical Technologies Department}, \orgname{National Physical Laboratory}, \orgaddress{\street{Hampton Road}, \city{Teddington}, \postcode{TW11 0LW}, \state{England}, \country{United Kingdom}}}



\abstract{
Quantum processors have the potential to revolutionise computing on a scale unseen since the development of semiconductor technology in the middle of the 20\textsuperscript{th} century. However, while there is now huge activity and investment in the field, there are a number of challenges that must be overcome before the technology can be fully realised. Of primary concern is the development of the classical technology required to interface with quantum systems, as we push towards a new era of high-performance, large-scale quantum computing.

In this review, we briefly discuss some of the main challenges facing the development of universally useful quantum computers and the different architectures being investigated. We are primarily concerned with cryogenic quantum systems. These systems are among the most mature quantum computing architectures to date, and are garnering a lot of both industrial and academic attention. 

We present and analyse the leading methods of interfacing with quantum processors, both now and for the next generation of larger, multi-qubit systems. Recent advancements in control cryoelectronics, both semiconducting and superconducting, are covered, while a view towards newer methods such as optical and wireless qubit interfaces are also presented.
}

\keywords{Quantum Computing, Classical Interface, Qubit, Superconducting, Cryo-CMOS, Wirelss, Optical, SFQ}



\maketitle

\section*{Introduction}\label{sec1}

Quantum computers have long been touted as the answer to ever-increasing demands for computing power, across a vast range of applications including cryptography, fundamental-physics research, and drug-discovery\cite{Steane1998, Spentzouris2020, Bonde2024}. The attention and focus on quantum computing (QC) is accelerating as industrial, academic, and government organisations invest significantly in developing the technology and realising its potential. While huge progress has been made in recent years\cite{MacQuarrie2020, Fedorov2022}, there are important obstacles and issues to overcome before true \textit{quantum advantage} is seen. A key area here concerns the quantum-classical interface needed to interact with and control the quantum processor\cite{Almudever2017}. Currently, within the noisy intermediate-scale quantum (NISQ) era\cite{Preskill2018}, quantum computers are limited to hundreds or potentially thousands of qubits, controlled by large numbers of cables and interconnects, and bulky, classical, electronic devices situated at various temperatures from cryogenic stages close to the qubit(s), right up to room temperature. The increasing space and cooling power requirements needed to sustain quantum-processing growth, in addition to the constant demand for fast, low-noise, high-fidelity electronics, makes tackling the issues surrounding the quantum-classical interface a high priority for the research field and industrial community. Here, we discuss the need to develop new or existing technologies and elucidate some of the current methods used. 


\section*{Advancing and scaling up the next generation of quantum computers}\label{sec2}

Scaling up the number of qubits comprising a quantum computer is a key aim of both industrial and academic researchers. The desire to move beyond the noisy intermediate-scale quantum (NISQ) era of computing is evident throughout the industry, as seen by IBM’s recent announcement to move towards a 100,000 physical-qubit system\cite{IBM2023IBM:Qubits}, and the investments made by governments all over the world towards developing quantum information processors that outperform current, classical high-performance computers\cite{UKDepartmentforScience2023UKStrategy, EuropeanCommission2022EU:Flagship}. Scaling quantum computers to their full potential involves overcoming a series of complex challenges. This necessitates a collaborative approach, uniting the expertise of scientists, engineers, and industry. Despite significant progress\cite{Gill2022, horowitz2019quantum}, there's still a considerable journey ahead. Central to this is engineering the appropriate classical interface for interacting with the quantum system. Among the primary challenges are:

\paragraph{Quantum decoherence} Quantum systems are extremely sensitive to environmental disturbances, leading to a loss of coherence known as decoherence. As the number of qubits increases, the probability of errors due to decoherence also rises. Developing error-correction techniques and improving qubit stability are crucial for scaling up quantum computers.\cite{Siddiqi2021}
	
\paragraph{Qubit interactions and connectivity} Complex quantum systems require qubits to interact with each other to perform complex computations. As the number of qubits grows, maintaining and controlling the interactions between them becomes increasingly difficult. Ensuring high connectivity and minimizing unwanted interactions are essential for scaling up quantum computers.

\paragraph{Qubit quality and manufacturing} Quantum computers require high-quality qubits with low error rates. Current qubit technologies, such as superconducting circuits and trapped ions, face challenges in achieving consistently high qubit quality at scale. Improving fabrication techniques, reducing manufacturing defects, and optimizing qubit performance are critical for large-scale QC.\cite{deLeon2021}

\paragraph{Hardware scalability} Building a large-scale quantum computer requires integrating a massive number of qubits and other components into a coherent and controllable system. The physical infrastructure, including cryogenic cooling, wiring, and control systems, must be designed to accommodate the increasing complexity and size of the quantum processor. This challenge will be a focus of discussion within this review.

\paragraph{Quantum algorithms and error correction} Scaling up quantum computers requires developing robust error-correction codes capable of detecting and correcting errors introduced during computation. Designing efficient fault-tolerant quantum algorithms that can tolerate errors while still providing meaningful results is an ongoing research challenge.\cite{Campbell2024}

\paragraph{Programming and software tools} As quantum computers scale up, designing and implementing quantum algorithms become more complex.\cite{Bharti2022} Developing user-friendly programming languages, software libraries, and simulation tools to facilitate not just quantum software development and optimization, but also the control systems for quantum hardware\cite{Shammah2024}, is crucial for wider adoption and efficient utilization of quantum computers.\cite{Cerezo2021}

\paragraph{Cost and resources} Quantum computers are currently expensive to build and operate. While progress has been made, scaling up QC requires significant investments in research, development, and infrastructure. Additionally, the demand for rare resources, such as high-quality qubit materials and specialized fabrication equipment, may pose supply chain challenges.

\section*{Qubit architectures}\label{sec3}
The focus of this review concerns superconducting qubits, one of the leading platforms in developing quantum processors\cite{Ezratty2023}. While other qubit architectures show promise in quantum information processing, many of these also require cryogenic conditions for operation, meaning the classical interfaces discussed herein are useful considerations across the field of QC. The main reasons for considering cryogenics are the need to establish the necessary physical operating conditions and for creating and maintaining a stable environment. Utilising operating temperatures significantly below a material's superconducting energy gap ($\Delta$) prevents random thermal excitations and enables Cooper pair formation for qubit realisation. External interference coupled to the system, such as heat or electromagnetic radiation, can cause qubits to decohere; cryogenic temperatures significantly reduce this environmental noise. In practice, this dictates the maximum operational temperature, especially considering that qubit frequencies generally lie in the low GHz range, which necessitates an optimal temperature in the low mK regime\cite{hotqp2018,Krinner2019}.

Superconducting qubits have been of interest for many years, and are currently among the most widely used and heavily invested qubit types in QC research\cite{Clarke2008SuperconductingBits, Wendin2017}. They are usually fabricated using multi-step additive and subtractive processes, involving lithographic patterning, metal deposition, wet/dry etching, and controlled oxidation of superconductor films. Reliable qubit fabrication at an industrial scale is still a major challenge, and an active area of research for many\cite{Verjauw2022,VanDamme2024}. Generally, materials such as aluminum or niobium are selected as the superconducting elements, and these are patterned onto silicon or sapphire substrates\cite{Kjaergaard2020SuperconductingPlay}.
 
Superconducting qubits show great promise for several reasons including their strong and relatively straightforward coupling to control signals, their relatively long and ever-increasing coherence times, and the clear path towards immediate scalability\cite{Oliver2013}. They must be operated at extremely low temperatures, typically around 10s of mK inside dilution refrigerators, to both ensure the devices are far below the critical temperature ($T_c$) of the component material, and to minimise the noise of the system\cite{Krantz2019AQubits, Kjaergaard2020SuperconductingPlay}.

Quantum dot qubits utilize semiconductor nanostructures, known as quantum dots, to trap and manipulate single electrons as qubits. They are typically fabricated using materials such as gallium arsenide or silicon, and operate at cryogenic temperatures, typically below 1 K\cite{Philips2022UniversalSilicon}. Spin qubits utilize the intrinsic spin of electrons or nuclei in solid-state systems, such as quantum dots or defects in crystals like nitrogen-vacancy (NV) centres in diamond. The cryogenic requirements for spin qubits can vary depending on the specific implementation. Some require operation at mK temperatures, while others can function at higher temperatures, around 1-4 K\cite{Burkard2023SemiconductorQubits}.


Trapped ion qubits use individual ions such as ytterbium or calcium. These ions are trapped and manipulated using electromagnetic fields in a vacuum chamber\cite{Bruzewicz2019Trapped-ionChallenges}. Photonic qubits use individual photons as qubits and manipulate them using various optical components such as beam splitters and detectors\cite{Slussarenko2019PhotonicReview, Ramakrishnan2023IntegratedReview}. Cryogenic temperatures are not required for trapped ion or photonic qubits but can be beneficial in improving their operation.\cite{Brandl2016, Pagano2019, Brown2021} For instance, some of the highest efficiency single-photon detectors, used to probe the state of a trapped ion or optical qubit, are based on superconducting materials that require Kelvin temperatures\cite{Todaro2021, Gyger2021, Zadeh2021}. The principles and proposed implementations of classical interfacing discussed here are relevant and potentially useful across many possible QC architectures.

\section*{Interfacing with quantum computers} \label{sec4}
To access the computing power of a quantum system, the quantum state of a qubit must be manipulated, and it must be possible to generate and analyse any qubit state using the control and readout electronics. A major barrier to scaling up quantum computers stems from the room temperature electronics and optics needed to control and readout these quantum operations, placing limits on the signal bandwidth, latency, thermal load, and signal-to-noise ratio in the system. As quantum computers move to interacting systems of thousands of qubits, more room temperature electronics are required. 
There is a need to exploit the potential of information and communication technologies (ICT) as an interface to develop scalable, efficient integrated control and readout architectures.

Before the different interfacing approaches are discussed, it is important to highlight the major constraint in integrating technology within a cryogenic system: limited cooling power. For example, while a cryo-CMOS processor operating at just 77 K is predicted to achieve 3.4 times higher performance\cite{Byun2021}, even cooling to this relatively high temperature can pose challenges. The 2022 IEEE international roadmap on \textit{Cryogenic Electronics and Quantum Information Processing} lays out many of the opportunities and challenges associated with integrating cryogenic systems.\cite{CE_QIP_RM2022} Of particular note for this work are the available cooling powers at various temperatures. At 4 K, for high powered cooling systems, there can be up to 1 kW of cooling power, however, at 20 mK there is only up to 30 $\mu$W of cooling power. These cooling limits are currently placing restrictions on not just quantum processor size, but also on the classical interfaces that can be placed within a cryostat.  


\subsection*{Conventional qubit control}
\begin{figure}[H]
  \centering
  \includegraphics[width=\textwidth]{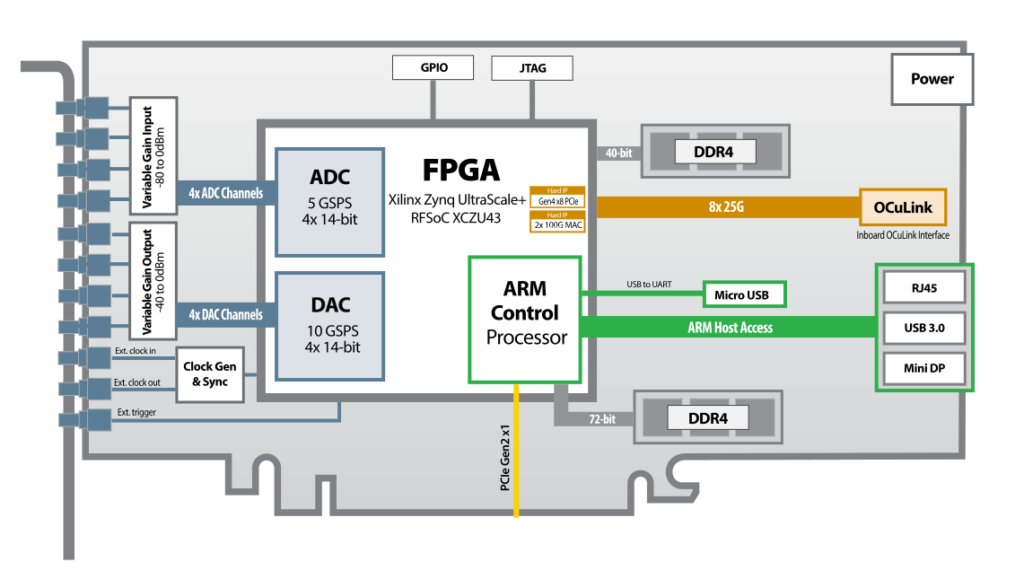}
    \caption{Structural diagram of an integrated measurement device where the analogue components are part of the integrated system on chip model architecture on a single integrated circuit board. The components are interacted through digital communication protocols through USB and Ethernet with a measurement PC. \cite{EntegraFPGA}}
    \label{fig:FPGA}
\end{figure}

\begin{figure}[H]
  \centering
  \includegraphics[width=\textwidth]{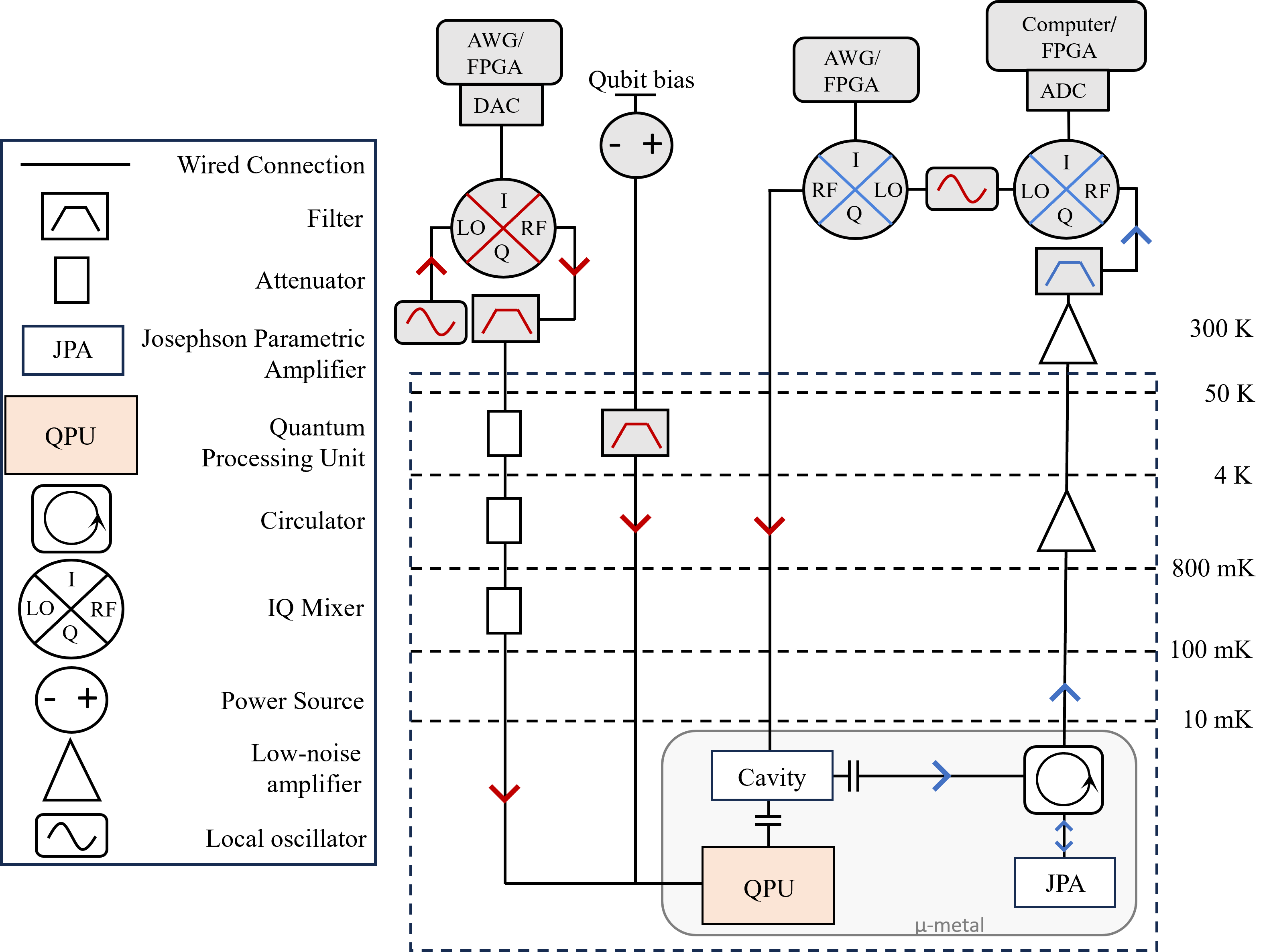}
    \caption{Schematic outlining conventional control and readout of a superconducting transmon qubit within a dilution refrigerator.}
    \label{fig:Conventional Control}
\end{figure}

A typical experimental setup involves two distinct microwave tones with similar properties, which pass through thermally anchored attenuator stages within a dilution refrigerator to suppress thermal noise before reaching the qubit sample. One of these signals is responsible for the control of the qubit state. By meticulously engineering various aspects of a microwave pulse, such as its amplitude, duration, and shape, it is possible to achieve any arbitrary state on the Bloch sphere within a timescale of tens of nanoseconds. An integrated measurement device is shown in Figure \ref{fig:FPGA}, while a simplified schematic for conventional control and readout can be seen in Figure \ref{fig:Conventional Control}. Advanced protocols with complex pulse sequences, such as the derivative removal by adiabatic gate or dynamical decoupling\cite{controlDRAG,controlDD}, are further employed to reduce the unwanted side effects of anharmonicity and environment noise in real qubit devices, aiming to improving the fidelity of the control operation. These pulses are typically generated using a heterodyne scheme, where low-frequency signals from an arbitrary waveform generator (AWG) define the shape and duration of the pulse. These signals are then mixed with a high-frequency carrier signal near the qubit's fundamental transition frequency to produce the desired control waveform.

Another microwave tone is used to probe the qubit state in a dispersive readout architecture. After interacting with the qubit through the resonator on-chip, the tone is amplified at several stages and then mixed with the source-reference tone and finally converted from analogue to digital and recorded for post-processing\cite{Krantz2019AQubits}.
Each step induces noise, reducing the signal-to-noise ratio and improving each step individually is a major source of research and effort, to enhance qubit control and readout.
Inside the dilution refrigerator, the measurement chain includes circulators and isolators, to suppress backaction signals to the qubit\cite{Pozar2011MicrowaveEngineering}, and amplification through quantum-limited Josephson parametric amplifiers (JPA) and high-electron mobility transistor amplifiers\cite{Caves1982QuantumAmplifiers, Clarke2008SuperconductingBits}.
Similar to the control tone, the probe tone performs a heterodyne/homodyne measurement, where a reference high-frequency carrier is mixed with a probe pulse from a digital to analogue converter (DAC) or AWG, to generate the readout tone by a process called upconversion\cite{Leonhardt1993RealisticDistributions, Wiseman2009QuantumControl}. After this, the transmission through signal is downconverted using the same reference carrier wave and recorded through an analogue to digital converter (ADC).
A major limitation stems from syncing all the individual drive electronics with an external clock, as they are usually generic and individually commercially sourced, this causes a delay ($\approx$1$\mu$s) which significantly affects gate operations.

A modern hybrid approach involves field programmable gate arrays (FPGAs) based on custom hardware that combines an entire rack of generic microwave hardware on a single PCB with integrated chips performing dedicated drive and readout tasks. These components can be digitally programmed on FPGA fabric, for example combining ADCs, DACs, filters, and attenuators on the same PCB, to record analogue signals and decipher qubit states in real time. This enhances gate performance significantly and enables fast feedback control of multiple qubits in one go\cite{Salathe2018Low-LatencyCommunication}. An example FPGA board is shown in figure \ref{fig:FPGA}.
Modern FPGA integrated circuits combine ADCs, DACs, and all other analogue RF and digital components onto a single die fabric, incorporating a \textit{System on Chip} philosophy to further reduce latency times between signal generation and readout to a few nanoseconds. 
The custom nature of these electronics enables these boards to be tailored to a quantum subsystem or generalised to perform an array of measurements from spectroscopy to multi-qubit control and sensor measurements\cite{Stefanazzi2022TheDetectors}. Multiple commercial vendors have utilized this opportunity to produce customized products for quantum measurements, with a varied range of control/readout frequency ranges and high bitrate (up to 14-bit) analogue signal capture\cite{QuantumMachines2023OPX+:Hardware}.

Overall, effectively reading out qubit systems at least requires high-speed classical electronics superseding qubit decay rates. Modern custom surface mount electronics coupled with FPGA/ASIC chips on board have been seen to be effective in multi-qubit control and readout. These classical electronics will need to keep improving, or new approaches will be required, as will be discussed here.

\section*{Scaling up} \label{sec5}

\subsection*{Conventional approach}
The scaling-up method currently being driven by large, industrial players in the field, Google and IBM for example, revolves around simply increasing the number of control and readout lines, proportional to the number of qubits in the system. This necessitates a direct increase in the size of the cryostat and the cooling power required, with projects such as IBM’s ‘Goldeneye’ developing a mK cryostat on a previously unseen scale\cite{IBM2022GoldeneyeSystem}. While this method has led to progress and may form the basis for the first generation of quantum computers, it is not a suitable long-term solution to the issue of scalability, as size and power requirements quickly become unreasonable for tens of thousands of qubits and above; this is particularly important amid a climate where power consumption is an important topic of research and discussion\cite{Chen2023AreEfficient, Fellous-Asiani2023OptimizingComputers}.


\subsection*{Cryo-CMOS Technologies}
In a conventional QC setup, control and readout electronics, operating at room temperature, are typically positioned several metres away from the qubits. Conversely, the qubits are located in a cryogenic environment, maintaining temperatures in the tens of mK range. The signal latency, stemming from the length of the cables connecting the electronics to the qubits, results in a delay of hundreds of nanoseconds. This substantial latency proves unsuitable for fast quantum feedback measurements\cite{Walter2017RapidQubits}, where the outcomes of single-shot qubit measurements immediately inform subsequent actions on the qubits. Reducing feedback latency is vital for minimising error rates in the feedback operation and enhancing the overall fidelity of quantum information processing. Moreover, as the quantum system scales up by increasing the number of qubits, challenges mount due to the growing number of cable requirements. To effectively tackle these issues, relocating electronics to cryogenic temperatures (cryoelectronics) presents a potential solution, alleviating signal latency and many cable concerns.

\begin{figure}[H]
  \centering
  \includegraphics[width=\textwidth]{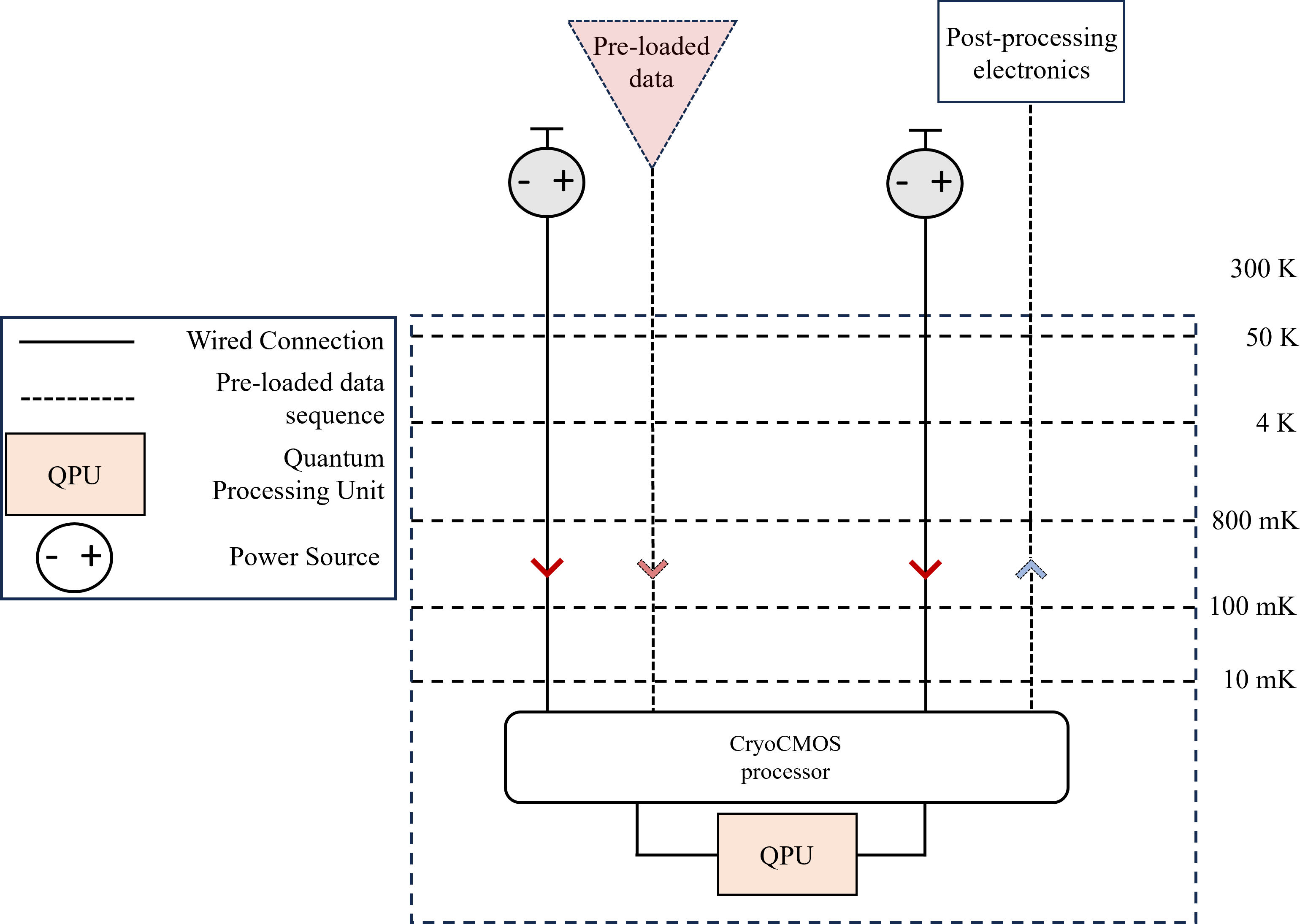}
    \caption{Block diagram outlining the principle elements of a cryo-CMOS controlled superconducting qubit in a dilution refrigerator.}
    \label{fig:CryoCMOS Control}
\end{figure}

Cryogenic electronics potentially empower rapid feedback mechanisms\cite{Walter2017RapidQubits}, allowing for quick error correction\cite{DiVincenzo2009Fault-tolerantQubits, Barends2014SuperconductingTolerance} due to reduced feedback latency, and ultimately enhancing the overall reliability of quantum computations, while preserving the integrity of quantum information.
In particular, cryogenic complementary metal-oxide semiconductor (cryo-CMOS) technologies have recently gathered significant attention from researchers due to their potential to revolutionize QC. Again, the distance between qubits at mK temperatures and the qubit readout and control circuitry operate at room temperature poses a significant problem that cryo-CMOS aims to overcome. \cite{Charbon_CryoCMOSforQC, Sebastiano2020Cryo-CMOSComputers, Sebastiano2017Cryo-CMOSInvited, Patra2018Cryo-CMOSApplications, Imroze2022PackagedApplications, Giagkoulovits2022CryoCMOSChallenges,8865458}. Cryo-CMOS technologies aim to address this issue by enabling the construction of custom-made readout and control circuitry that is designed to operate at cryogenic temperatures. A potential QC - cryo-CMOS architecture is presneted in Figure \ref{fig:CryoCMOS Control}. However, specific challenges are associated with building such circuits, including power budget, noise, and bandwidth constraints\cite{Mehrpoo2019BenefitsComputers, Charbon2021CryogenicProcessors}.  

Two major technologies are currently being used in cryoelectronics QC hardware: bulk CMOS and fully depleted silicon-on-insulator (FDSOI)\cite{Nayak2021FullTemperatures}. Bulk CMOS technology, as in conventional CMOS technologies adapted to operate at cryogenic temperatures, offers low complexity and cost-effectiveness compared to FDSOI but poses several performance-related challenges when operated at cryogenic temperatures, such as carrier freeze-out, increased leakage current, kink effect and hot carrier degradation\cite{Zhang2021HotK, Beckers2018CharacterizationK, Lu2020CharacterizationTemperature}. FDSOI technology, on the other hand, is more sophisticated and offers greater benefits for cryogenic applications, including reduced leakage currents, low power consumption, and back biasing to counter threshold voltage shift at cryogenic temperatures\cite{LeGuevel2020Low-powerDevices, Yang2020AFDSOI}. 


A critical aspect of advancing cryogenic CMOS technologies is the development of a cryogenic process design kit (PDK). This task involves characterizing devices at cryogenic temperatures as a first step. Room temperature data is verified from the foundry PDK if it falls within the process corners. The data is then calibrated at the desired cryogenic temperature using a technology computer aided design tool through reverse engineering. Compact model parameters are extracted, and the existing room temperature PDK is tuned to create a cryogenic PDK for that device.

While cryo-CMOS control potentially offers significant advantages for scaling QC, it does come with several limitations. One important constraint is the same issue conventional control has to tend with: when dealing with a large number of qubits, the number of control lines must still increase linearly\cite{Mehrpoo2019BenefitsComputers, Charbon2021CryogenicProcessors}. One way to somewhat alleviate this issue would be to pre-load a data sequence into the cryo-CMOS processor in the cryostat. This would mean that only a power source would need to be connected to the cryo-CMOS processor through the cryostat, and signals would not need to be sent in to the cryostat, from instruments outside the fridge, during measurements. However, there would still need to be many connections to the quantum processor. While these connections could be made using superconducting materials, below 4K, to minimise the heat conduction between stages and electrical losses, the cryo-CMOS control circuits will still encounter issues with significant heat dissipation. This issue can potentially surpass the cooling capacity of the cryostat, resulting in an undesired increase in the qubit's temperature, disrupting all aspects of its operation.\cite{VanDijk_CryoCMOS_AnalogSignals}. 

A review of the current state of research and development shows that prominent technology companies such as Google\cite{Yoo202334.2Unit-Cell}, Intel\cite{Park2021AQubits}, and IBM\cite{Chakraborty2022ATechnology} have played pivotal roles in advancing the integration of cryoelectronics to within the proximity of the qubit stage (4K and below) by application specific integrated circuits (ASIC) using cryo-CMOS technologies. Indeed, the first European \textit{Chips Joint Undertaking} project ARCTIC (Advanced Research on Cryogenic Technologies for Innovative Computing)\cite{ChipsJU_ARCTIC} has started to establish a European supply chain for cryogenic quantum technologies, including cryogenic photonics, microelectronics, and cryo-microsystems to foster the low-cost, high-volume manufacturing of superconducting and semiconductor-based quantum chips in the EU.

What stands out as a remarkable achievement across these research efforts is the demonstration of practical control and readout architectures that function at cryogenic temperatures and exhibit high reliability and effectiveness, all with limited power dissipation. This concerted effort among these industry leaders underscores the significant strides in pursuing QC and quantum information processing technologies, with cryoelectronics serving as a critical enabler for advancing these QC applications.
Overall, cryo-CMOS technology offers a potential solution to the scalability of quantum computers, provided a few implementation challenges are addressed. Initially, substantial effort is required to develop reliable cryogenic models that accurately replicate the cryogenic performance of CMOS circuits, before bespoke devices can be designed and fabrication processes can be initialised and industrialised.

\subsection*{Single flux quantum}
Long-standing single flux quantum (SFQ) technology has recently been demonstrated to manipulate and control qubits in superconducting circuits\cite{Leonard2019DigitalQubit,Liu2023SingleModule}. This is a digital approach that utilizes the properties of superconducting materials, in the form of Josephson junctions, and the quantization of magnetic flux to encode and process quantum information\cite{Likharev1991RSFQSystems, Bunyk2001RSFQDevices}. An SFQ pulse is produced when magnetic flux through a superconducting circuit containing an overdamped Josephson junction changes by one flux quantum, $\Phi_0$, as a result of the junction switching, which represents a 2$\pi$ phase jump. SFQ pulses have a quantized area 

\begin{equation}
    \int V(t) dt = \Phi_0 \approx 2.07 \cdot 10^{-15} \text{ Wb} = 2.07 \text{ mV.ps} \label{SFQarea}
\end{equation}
\\
Pulses can have a time interval as short as 1 ps, with an amplitude in the mV range, depending on the characteristic voltage of the circuit. 
In rapid single flux quantum (RSFQ) logic, one of the oldest and most developed logic families, propagating voltage pulses represent classical bits of information: the presence or absence of a voltage pulse across a Josephson Junction in the circuit constitutes a classical 1 or 0 binary bit. Traditionally, RSFQ devices have been used in ultra-fast electronics, including ADCs, RF transceivers, microprocessors, as it has long been possible to drive them to very high frequencies, with simple circuits shown to run at clock speeds of 770 GHz\cite{Chen1999RapidGHz}.

\begin{figure}
 \centering
 \begin{subfigure}[b]{0.6\textwidth}
  \includegraphics[width=\textwidth]{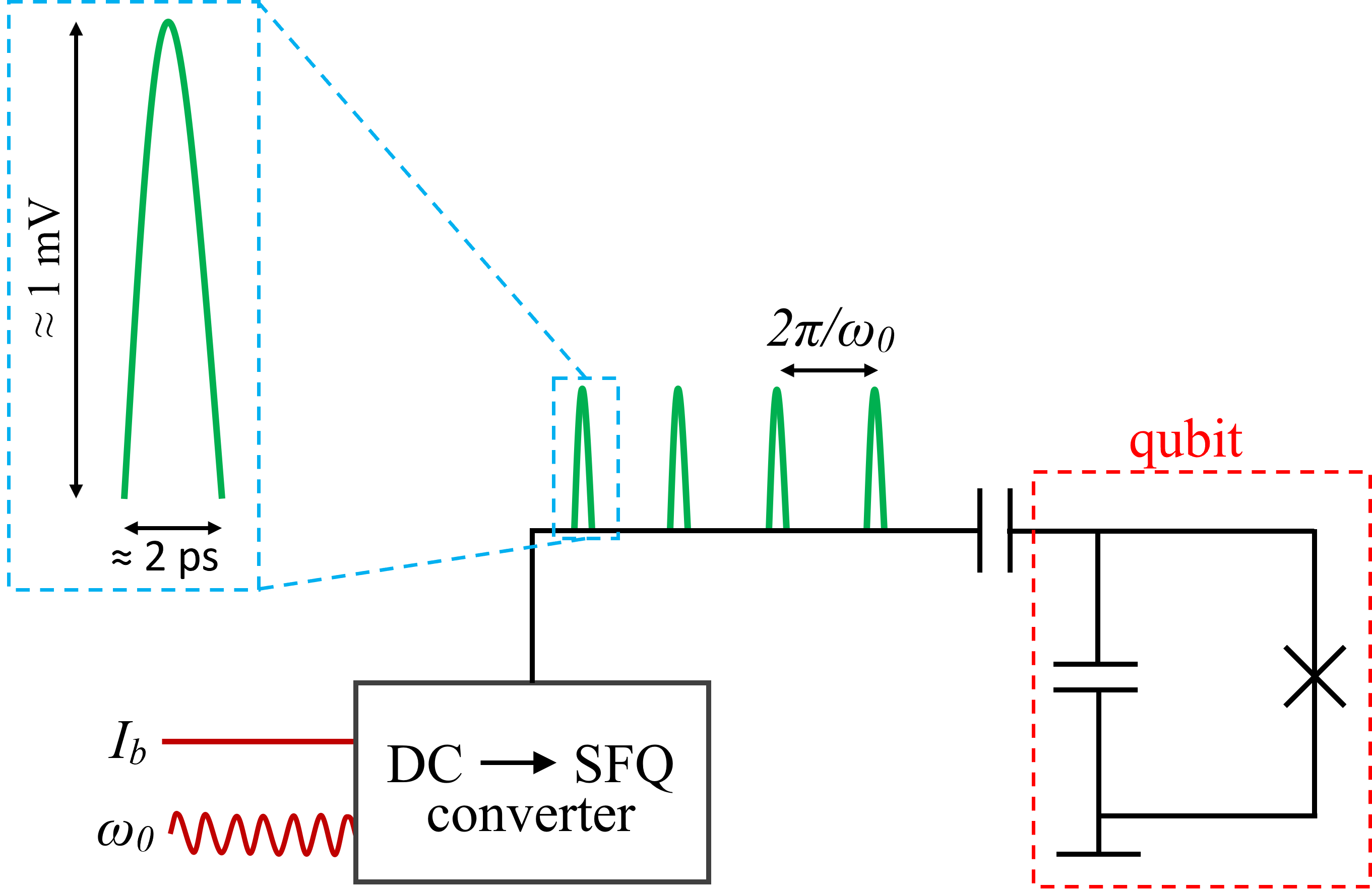}
  \caption{}
 \end{subfigure}
 \hfill
 \begin{subfigure}[b]{0.3\textwidth}
 \includegraphics[width=\textwidth]{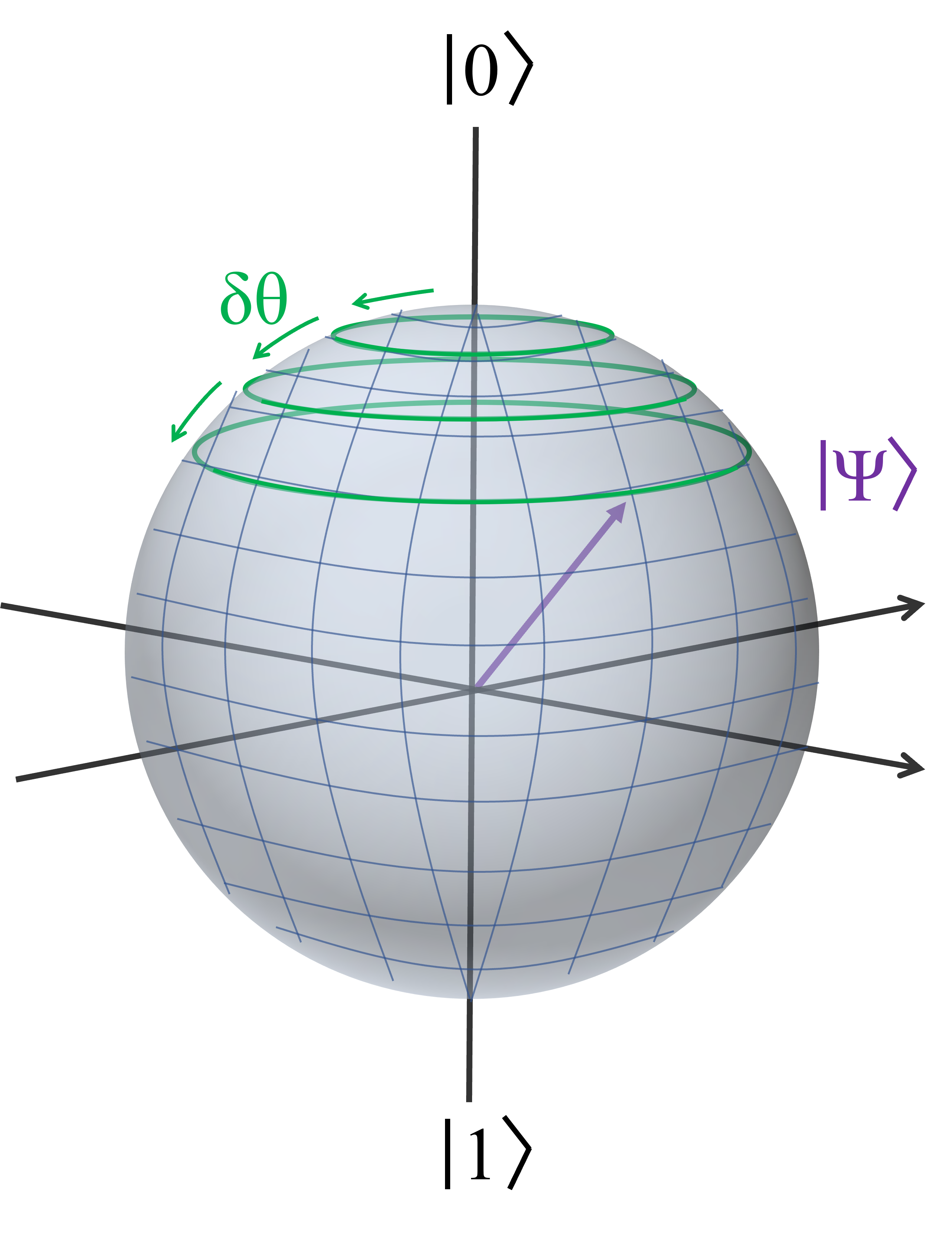}
  \caption{}
 \end{subfigure}
\caption{Schematic for coherent qubit control based on a train of SFQ pulses. (a) In SFQ operation, a voltage pulse, with time integral quantized to $\Phi_0$, is induced by a phase difference across a Josephson junction. A microwave signal of frequency $\omega_d$ is used to trigger a dc-SFQ converter, biased by a current $I_b$ producing a train of pulses at $\frac{2\pi}{\omega_0}$ intervals, which is capacitively coupled to a transmon qubit. (b) Each SFQ pulse induces an incremental rotation of the qubit's state vector on the Bloch sphere.}
    \label{fig:SFQ Function}
\end{figure}

An individual SFQ pulse produces a broadband excitation, so is incompatible with coherent manipulation of qubit systems, as specific qubit transitions can not be isolated and excited\cite{Barbosa2024}. Instead, the theory behind SFQ control of quantum systems relies on driving them using a train of SFQ pulses. The qubit is coherently excited using a pulse-to-pulse separation matched to its period\cite{McDermott2014AccuratePulses, Geng2023QubitCircuits}. 
As shown in Figure \ref{fig:SFQ Function}, an external microwave tone is used to trigger a DC-SFQ converter, which generates the pulse train. The pulse train is then coupled capacitively to a transmon qubit, and each SFQ pulse induces an incremental rotation of the qubit state vector about the Bloch sphere. In this way, coherent and accurate qubit control can be achieved, with a predicted fidelity of $>$99.99\%\cite{Li2019Hardware-EfficientSequences}. This has since theoretically been extended to multi-gate systems with promising results for maintaining high-fidelity control\cite{Jokar2021PracticalGates}.
Experimentally, this approach has now been realised, with initial gate fidelities of $\approx$95\%\cite{Leonard2019DigitalQubit}, which have since been improved by redesigning the layout of the SFQ-qubit system using a multichip module\cite{Liu2023SingleModule}. The process of scaling up the number of qubits has now begun, with Bernhardt \textit{et al.} reporting the first multi-qubit system integrating SFQ technology.Here, they utilise digital demultiplexing, breaking the linear scaling of control lines to number of qubits, showing promise for heat-load and space savings in cryogenic QC\cite{bernhardt2025quantumcomputercontrolledsuperconducting}.

SFQ-based qubit control is therefore a possible and promising avenue for qubit systems, but SFQ logic has also been touted for use in qubit readout. As discussed, superconducting qubits are generally measured using heterodyne detection of a coupled resonator, requiring significant equipment overhead throughout the cryostat and at room temperature for each qubit. A method has been presented by Opremcak et al.\cite{Opremcak2018MeasurementCounter} and Howington et al.\cite{Howington2019InterfacingReadout}, whereby a qubit state is mapped onto the photon occupation in a microwave cavity, after which the photon is detected using a Josephson photomultiplier (JPM), which effectively measures the qubit state and stores the result in a circulating current. An underdamped Josephson transmission line (JTL) coupled to the JPM delays or accelerates fluxons traveling along the JTL, depending on the circulating current in the JPM. The deviation in fluxon arrival time is converted to an SFQ logic signal, delivering digital qubit readout.

Another readout method has also been recently proposed, based on amplitude/phase discrimination of a microwave tone, whereby the phase of a coherent microwave tone can be probed between two distinct phase values by a device called Josephson digital phase detector (JDPD)\cite{DiPalma2023DiscriminatingCircuit}. The JPDP can then be coupled to a qubit readout resonator and be able to distinguish between the two qubit states, which have a fixed phase difference due to the dispersive readout scheme. Interfacing the JPDP with SFQ electronics would be straightforward due to the nature of the device, creating an alternative way of fully digital readout of superconducting qubits.

\begin{figure}[H]
  \centering
  \includegraphics[width=\textwidth]{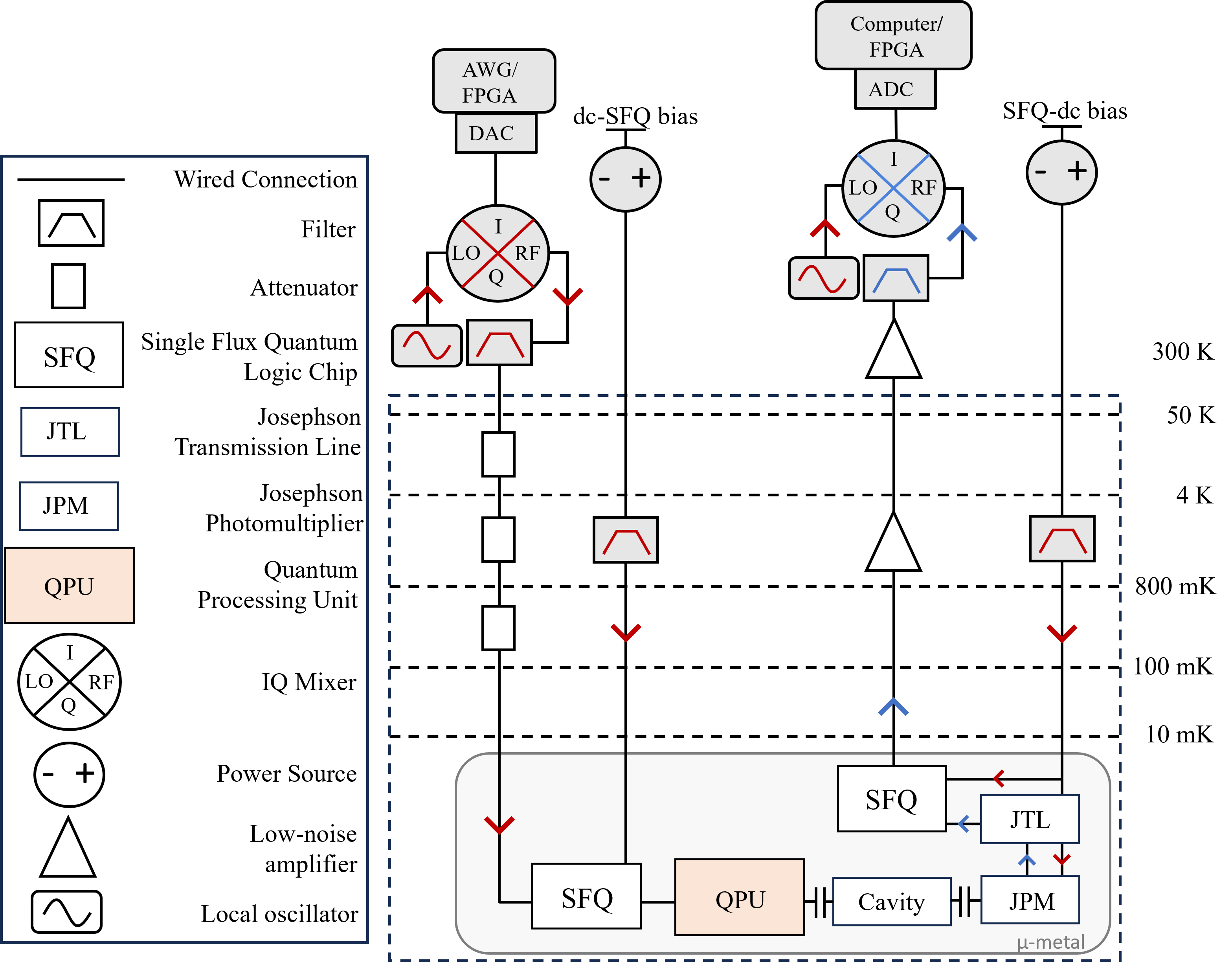}
    \caption{Schematic showing the main elements of a control and readout architecture, for interfacing with superconducting qubits in a dilution refrigerator, using SFQ technology.}
    \label{fig:SFQ Control}
\end{figure}

An example of an SFQ-qubit architecture is presented in Figure \ref{fig:SFQ Control}. There are of course some drawbacks with implementing SFQ control and readout of qubit systems. A significant issue is the implementation and integration of SFQ devices into the commercially available dilution refrigerators commonly used for superconducting qubit research. The low-noise requirements and sensitivity of the devices to magnetic fields and trapped flux means that manageable, but not-insignificant modifications are required to adapt the system from conventional qubit control architecture. Additionally, quasiparticle poisoning was seen to be an issue with early implementations of SFQ control. This has recently been improved significantly through installing the SFQ device at the 3K cryostat stage\cite{Howe2022DigitalK} or implementing a flipchip module architecture\cite{Liu2023SingleModule}. A useful overview of SFQ-quantum interface technologies can be found at \cite{McDermott2018Quantum-classicalLogic}.


Overall, SFQ technology is particularly well-suited for use in quantum computers due to its high speed, low energy consumption, and compatibility with existing superconducting circuitry. Modern energy efficient implementations of RSFQ (ERSFQ) devices can display a bit-switch energy of around $2\times10^{-19}$ J with zero static power dissipation. Additionally, adiabatic quantum-flux-parametron (AQFP) logic has emerged as a promising candidate for qubit control, offering even lower power dissipation due to its inherently reversible computational nature\cite{aqfp24}.
Even factoring in the energy cost of cryocooling the devices to below T\textsubscript{c}, SFQ-based systems are projected to be more efficient than standard CMOS-based technologies for large scale computing\cite{Fellous-Asiani2023OptimizingComputers}.

\subsection*{Optical control and readout}
\begin{figure}[t]
  \centering
  \includegraphics[width=0.9\textwidth]{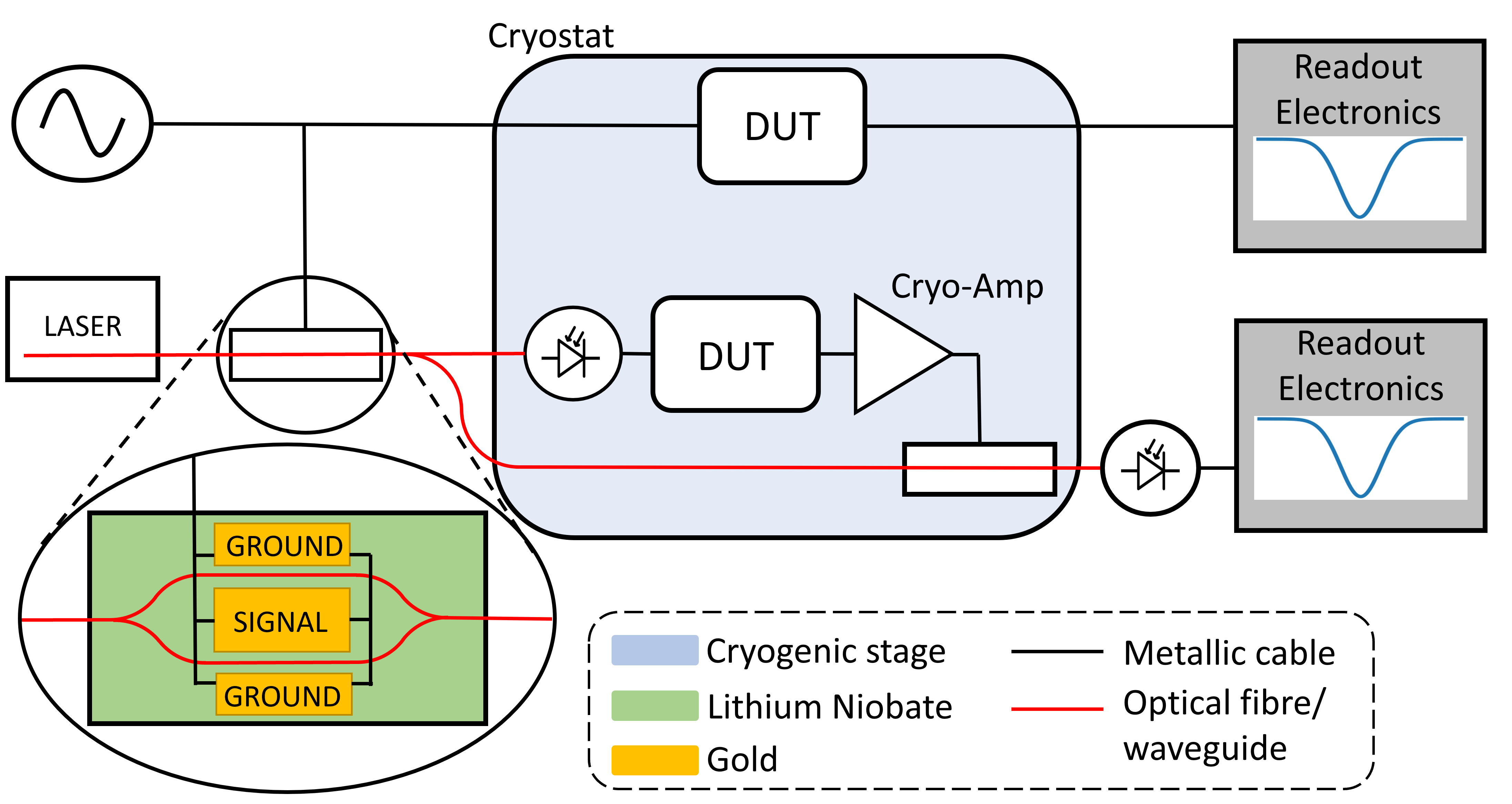}
    \caption{A potential electro-optical control and readout method for superconducting devices based on work presented in \cite{Youssefi2021ADevices} and \cite{Lecocq2021ControlLink}. The top circuit shows a simplified model of conventional readout techniques using metallic cables, with the bottom diagram showing a method utilising electro-optical components for transduction and optical fibres for signal transmission to cryogenic temperatures. Inset shows the design of an intensity modulator, the upper and lower waveguides cause an opposing increase and decrease in optical path length due to the opposite electric field in each arm, resulting in constructive and destructive interference at the output of the device.}
    \label{fig:Optical Interface}
\end{figure}

The potential for electro-optical technologies to be integrated with quantum devices is becoming an increasingly popular point of discussion. This is an area of research in which long-established techniques, in use globally for fibre-based information transmission, can be applied to the fields of QC and quantum sensing. The primary issue that can be addressed with electro-optical techniques is that of heat dissipation from coaxial cabling within the cryostat. The relevant technologies for enabling this technique will be discussed here alongside some examples of their use in cryogenic signal transmission applications; an example schematic can be seen in Figure \ref{fig:Optical Interface}.
The methods required for electro-optical signal transmission can be split into two main areas: encoding electrical signals onto the optical domain, and recovery of the electrical signals. Encoding is generally carried out by an electro-optical modulator (EOM), which takes advantage of the Pockels effect in dielectric crystals. Pockels materials have refractive indices which are proportional to an applied electric field, this can be approximated to first order by equation \ref{Pockels}.

\begin{equation}
n_{eo} \approx n_x \Bigl( 1 - \frac{1}{2}\gamma n_x^2 |\vec{E}| \Bigr) \label{Pockels}
\end{equation}
\\
Where $n$\textsubscript{x} and $n$\textsubscript{eo} are refractive index with and without an applied electric field, $\gamma$ is the Pockels coefficient, and $|\vec{E}|$ is the electric field in the material. This property can be exploited to control the optical path length of light propagating through the electro-optic crystal and in combination with optical waveguides, devices can be realised which control the phase, intensity, and polarisation of coherent light\cite{Sinatkas2021Electro-opticPhotonics}.
There are a range of metrics for quantifying EOM performance; when considering an EOM for use in cryogenic measurements, the location of the modulator will determine which metrics to prioritise. If an intensity modulator is to be used for encoding information on optical signals at room temperature, then a high extinction ratio (ER) is crucial. ER is the ratio between output in the modulators ‘on’ state and ‘off’ state; a high ER is desirable as it minimises the laser power required to effectively transmit information. In practice, lithium niobate intensity modulators have been tested, with ERs upwards of 20dB\cite{Weigel2018BondedBandwidth}.

If the EOM is to be operated at cryogenic temperatures, other metrics must be taken into consideration, such as insertion losses, power consumption, and modulation efficiency. Insertion losses and power consumption both relate to the heat dissipated by the EOM in the cryogenic stage. Coupling of the optical fibre mode to the waveguide mode as well as dielectric losses will determine insertion losses, and the issue of coupling is complicated by the thermal stresses of cooling down to mK temperatures. It is not clear which approach to coupling works best but the work in Ref. \cite{Youssefi2021ADevices} estimates that transmission of a device with glue bonds was reduced by 25\% through the cooldown process. Power consumption is strongly related to electrode resistance, which can be greatly reduced by employing superconducting electrodes\cite{Yoshida1999AElectrodes}.

Recovering an electrical signal from intensity modulated light is most simply carried out directly by a photodiode. If the optical signal is phase modulated, then an interferometer such as a Mach-Zender interferometer can be employed. Photodiodes are capable of detecting 1550 nm light at mK temperatures and can have a bandwidth of 10s of GHz, which is suitable for current superconducting qubit measurements. Photodiodes have been demonstrated to provide shot-noise-limited performance in an electro-optical control scheme for qubits\cite{Lecocq2021ControlLink}. One drawback of commercially available photodiodes is the ubiquity of 50-Ohm outputs lines. A photodiode with higher output impedance could achieve increased conversion gain, thus decreasing the laser power required.
\begin{figure}[t]
  \centering
  \includegraphics[width=\textwidth]{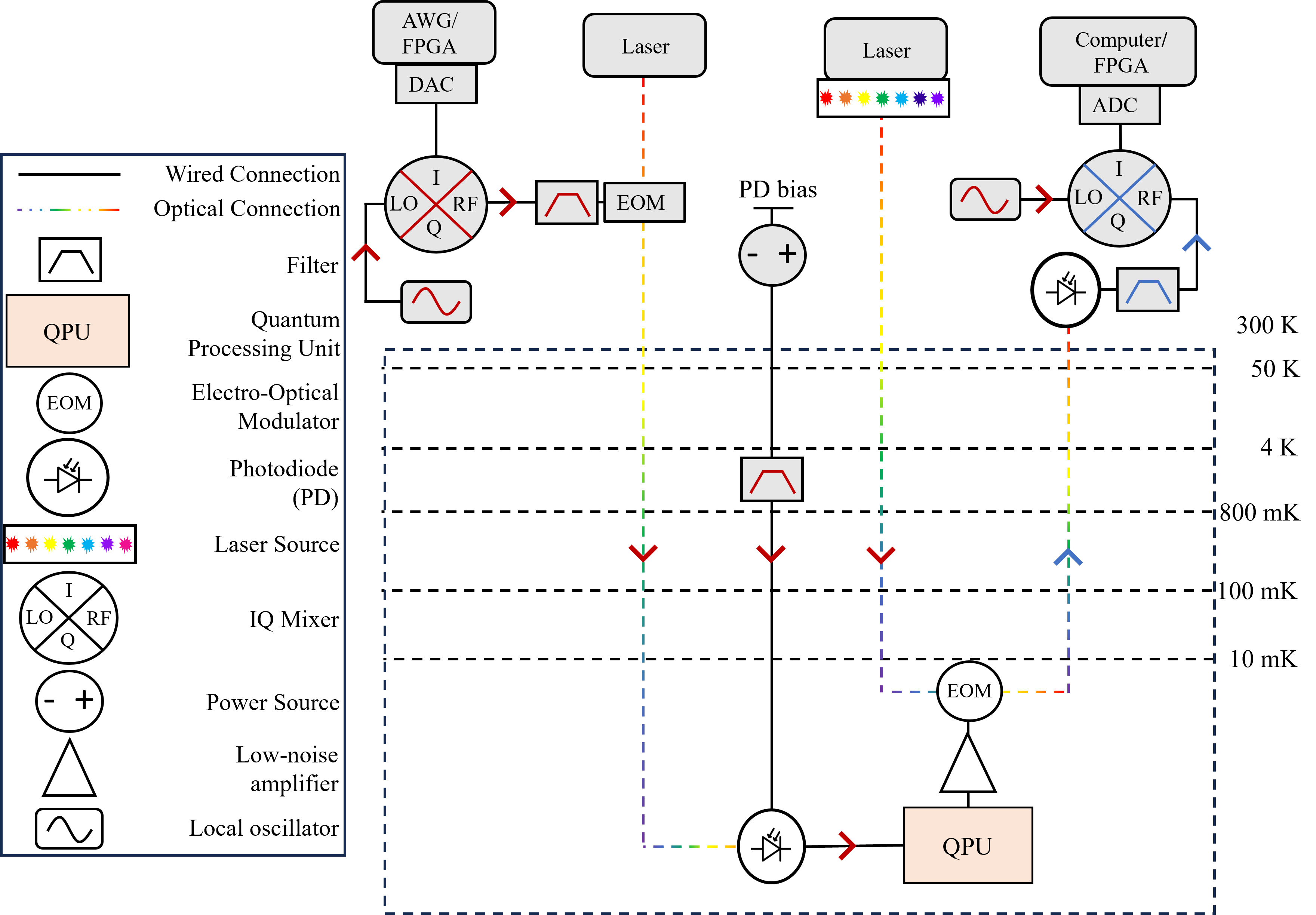}
    \caption{Schematic showing the main elements of a proposed architecture for optically interfacing with superconducting qubits, within a dilution refrigerator.}
    \label{fig:Optical Control}
\end{figure}

The idea to use electro-optics to replace current cryogenic signals transmission techniques has gained some attention. In \cite{Lecocq2021ControlLink}, the authors demonstrated low-noise qubit control and readout by replacing coaxial input lines with an electro-optical scheme. In \cite{Youssefi2021ADevices}, a superconducting electromechanical device was measured using a phase modulator at cryogenic temperatures to encode the signal. There has been some attention in the field of quantum sensing, with work carried out to control and readout superconducting single-photon detectors using electro-optical methods\cite{deCea2020PhotonicDetectors, Thiele2023AllInterface}. Recently, full readout of superconducting qubits has been shown by Arnold \textit{et al.}, where they use radio-over-fibre down to mK temperatures, in an optical heterodyne detection scheme\cite{Arnold2025}. In this work, they show simultaneous upconversion and downconversion between microwave and optical frequencies, removing the requirement for any active or passive cryogenic microwave equipment. A low optical coupling efficiency means heat from the parametric pump being dumped at the mixing chamber, leading to temperature increase and some degradation of the qubit coherence. However, this set-up has huge potential in reducing the heat-load and space needed inside the DR, and lays out a path to meet the next round of engineering challenges to achieve qubit scale-up.

\subsection*{Wireless control and readout}
\begin{figure}[H]
  \centering
  \includegraphics[width=0.9\textwidth]{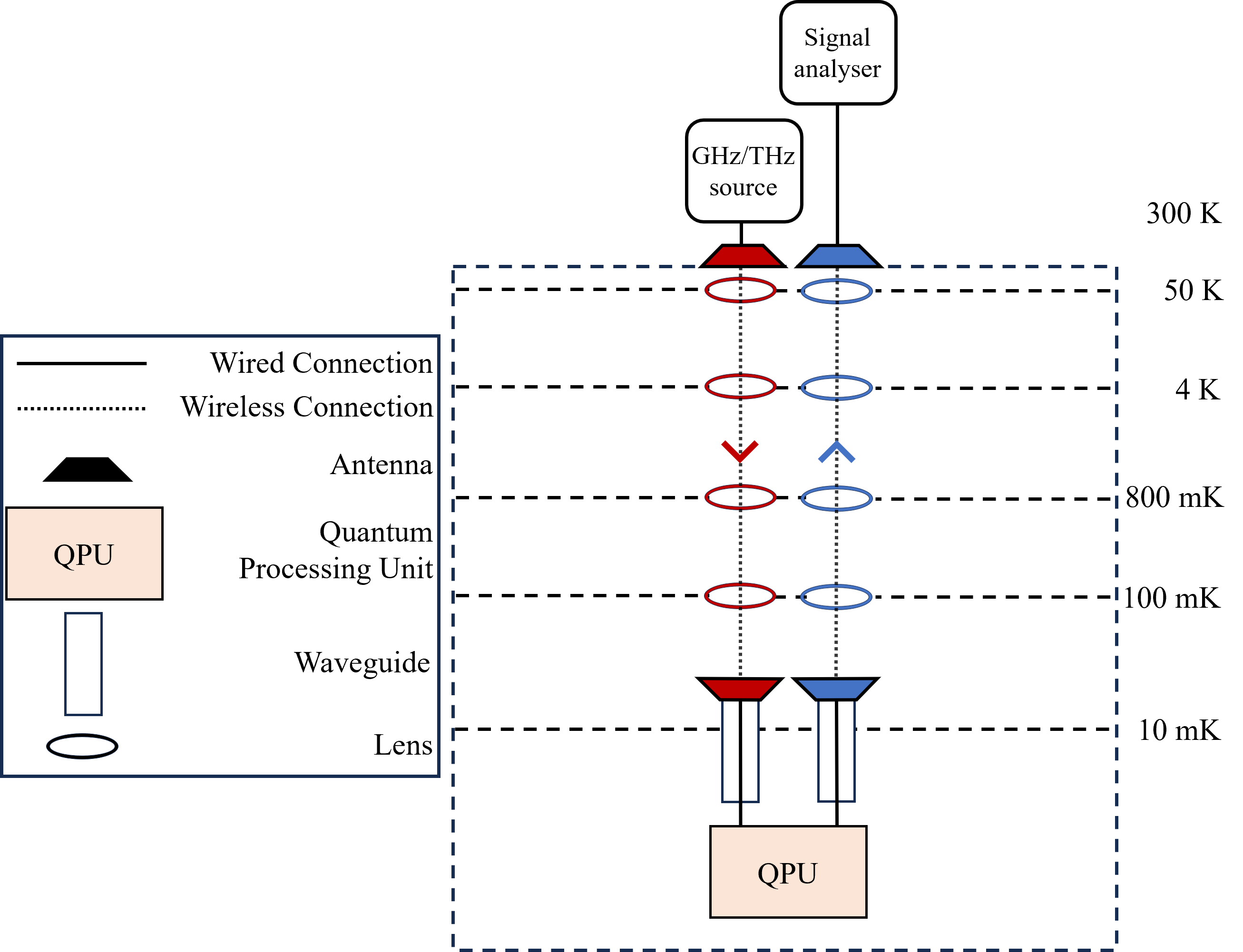}
    \caption{Schematic showing an overview of the proposed wireless qubit control and readout architecture within a dilution refrigerator.}
    \label{fig:Wireless Control}
\end{figure}

An alternative and recently studied solution is to implement wireless technology into QC systems\cite{Al-Moathin2023CharacterizationApplications, Kazim2023WirelessApplications, Wang202334.1Interface, Anders2023CMOSSciences, Alarcon2023ScalableNetwork-in-package}. Conventional wired connections for addressing and control of qubits can be directly replaced with wireless links\cite{Al-Moathin2023CharacterizationApplications, Kazim2023WirelessApplications}. Figure \ref{fig:Wireless Control} illustrates a possible setup with separate control path (downlink, or DL) and readout path (uplink, or UL) inside the QC system. A reflection mode with DL and UL sharing the same optical path is also possible. The wireless links include transceivers and multiplexing units at both room temperature (RT) outside of the refrigerator and the mK stage inside the refrigerator, as well as beam control components including filters, and collimators or lenses linking the RT and mK transceivers. This technique has potentially many advantages over the state-of-the-art wired solutions:
\begin{enumerate}
    \item All bulky and energy-hungry but highly sophisticated and relatively mature control and digital electronics remain at RT, making the cryostats much smaller, potentially leading to new and more compact designs where excitations and readouts via side walls are possible leaving the overall size of a QC only determined by the cooling system;
    \item Wireless links can accommodate thousands or more channels (microwave tones with spacing of a few MHz subject to qubit technologies) co-existing and therefore qubit upscaling is no longer limited by the size of the refrigerator. In addition, many established wireless technologies such as channel coding and signal modulation can be implemented to improve control, robustness, resilience, and security;
    \item Wireless links don’t need a transmission medium, vacuum inside of the refrigerator makes signals propagate 33\% faster than those in the coaxial cables. Thus, the latency between RT electronics and the qubits can be reduced by at least 33\% and potentially more with shorter paths which is crucial for high fidelity in many qubit technologies;
    \item Thermal imbalance between the inner and outer conductors of the coaxial cables requires additional solutions which increases overhead of QC systems. More importantly, the wired solutions bring excessive heat to the chamber, increasing the burden on the cryogenic cooling especially when the number of qubits rises to hundreds or more; on the contrary, the wireless solution has no such problems as there is no transmission loss in vacuum and the transceivers on the mK stage do not contain additional active devices but passives such as antennas and matching circuits as well as qubit circuits. Since the passives could be realised by superconductors that have negligible ohmic loss at that stage, no extra power will be required apart from what is required by the qubits. Qubits’ receiving antennae and multiplexing circuits could also be replaced by direct sensing with gate resonators in future developments.   
    \item There is much room for scalability to facilitate higher frequency signals. With millimetre wave frequencies, superconducting qubits could be operated at elevated temperatures, relaxing cooling power requirements, and higher processing frequencies. An example of this higher frequency operation has recently been shown by Anferov \textit{et al.} for superconducting qubits\cite{anferov2024millimeterwavesuperconductingqubit}.
\end{enumerate}

 
 

However, there are still many challenges including unknown signal propagation and scattering within the chambers, channel interferences and crosstalk, stray field associated noise on qubit performance, thermal control at the interfaces between ambient and the refrigerator and between chambers inside the refrigerator, realising transceiver and multiplexing circuits for mK operations as well as numerous unforeseen challenges from the diverse qubit technology. New cryogenic compatible microwave free space components are required, for example antennas \cite{Al-Moathin2023CharacterizationApplications} and lenses \cite{Kazim2023WirelessApplications} etc.  

An alternative wireless approach was proposed at MIT, which utilises backscattering technology. In this approach, a CMOS transceiver chip is placed inside the fridge to receive and transmit data. Terahertz waves generated outside the refrigerator are beamed in through a glass window. Data encoded onto these waves can be received by the chip. This chip also acts as a mirror, delivering data from the qubits on the terahertz waves it reflects to their source. This reflection process also bounces back much of the power sent into the fridge, so the process generates only a minimal amount of heat. The contactless communication system consumes up to 10 times less power than systems with metal cables\cite{Wang202334.1Interface, Anders2023CMOSSciences}. One of the challenges with this technology is that photon energy at THz is close to room temperature, and addressing how to reduce the impact of ambient temperature noise on these qubits is crucial. More recently, continuing the work, Wang \textit{et al.} report the use of a wireless terahertz cryogenic interconnect combined with cryo-CMOS technology.\cite{Wang2025} Wideband transceivers with a carrier frequency of 260 GHz, a hot-to-cold ingress based on passive cold field-effect transistor terahertz detector, and a cold-to-hot egress using ultralow-power backscatter modulation at the cold reservoir define this system, where they show that the ratio of information transfer to thermal heat transfer approaches the fundamental limit of what an equivalent optical interconnect can achieve. Again, they set out a path to the next stage of engineering problems needing to be solved to allow wireless scaling of multi-qubit systems.


\subsection*{Power and scalability considerations}
\renewcommand{\arraystretch}{1.2}
\begin{table}[t]
    \centering
    \setlength{\tabcolsep}{5pt}
    \begin{tabular}{c|c|c|c|c|c|} 
        Technology & \makecell{\# Drive and\\ readout lines} & \makecell{\# Flux lines} & \makecell{Controller-qubit\\distance}  & \makecell{Power consumption\\ per physical qubit} \\ \hline
        Standard Microwave & 5-200/100 qubits\footnotemark[1] & 1/qubit & RT-mK & $>$1 W \cite{8865458} \\
        Cryo-CMOS & 5-200/100 qubits\footnotemark[1] &  1/qubit & 4K-mK & 2-30 mW \cite{8865458,PRXQuantum.4.040319}\\
        SFQ logic& None\footnotemark[2]& None\footnotemark[2] & mK-mK & $<$1 nW \cite{McDermott2018Quantum-classicalLogic,Liu2023SingleModule}\\
        Optical& None\footnotemark[3] & 1/qubit & RT-mK &  \makecell{$<$10 nW (Optical-RF)\cite{Youssefi2021ADevices}\\$<$10 \textmu W (RF- Optical)\\\cite{Lecocq2021ControlLink,Shen24photonicSFQ}}\\
        Wireless & None  & 0.5/qubit\footnotemark[4] & RT-mK & $<$1 nW \\
    \end{tabular}
    \footnotetext[1]{Assuming an optimal control and readout multiplexing scheme \cite{PRXQuantum.4.040319}, and a worst case of a single control and readout line per qubit}
    \footnotetext[2]{Assuming an integrated SFQ-qubit approach with capacitive and inductive couplings between the two elements \cite{Leonard2019DigitalQubit,Liu2023SingleModule}}
    \footnotetext[3]{Drive and readout lines replaced with optical fibers.}
    \footnotetext[4]{Provided polarisation and (or) modulation schemes are applied}
    \caption{Comparison of classical interfaces with respect to scalability and power consumption.}
    \label{tab:table1}
\end{table}
To evaluate the potential of different technologies for use in future large-scale systems, it is essential to consider several key metrics. Table \ref{tab:table1} highlights the most relevant metrics for scaling superconducting quantum computing platforms beyond the NISQ era. These include the number of coaxial lines, such as drive, readout, and flux lines, connecting the controller units to the physical qubits inside the cryostat, the physical separation between these components across the stages of a conventional dilution refrigerator, and an estimation of the power consumption required to operate a single physical qubit effectively.

The number of physical wiring lines within the cryostat is a critical consideration as the number of qubits scales up. The internal volume and cooling power of a cryostat place a hard limit on the number of wires, particularly coaxials, that can be used. Subsequently, any future control system based on physical wiring will probably require multiplexing architectures \cite{freqmultiplex2012,tdmultiplex2023}. The most common of these is a frequency multiplexing system, where a single line is used to encode signals with multiple frequencies, each targeting a certain component of the qubit array. Based on Ref. \cite{PRXQuantum.4.040319}, the values shown in Table \ref{tab:table1} for conventional and cryo-CMOS technologies present an optimistic multiplexed control and readout architecture. This features 100:1 readout resonator multiplexing, corresponding to a 10 MHz separation within a 1 GHz bandwidth, and a 25:1 multiplexing of control lines, corresponding to a 50 MHz distancing within a 1 GHz bandwidth.  In practical near-term quantum systems, the best readout multiplexing ratios decrease to 8:1, as seen in IBM's quantum processors \cite{IBMQuantumRoadmap}. This limitation arises from the constrained bandwidth of quantum-limited parametric amplifiers, which restrict the number of frequency-multiplexed signals that can be amplified simultaneously. Additionally, each qubit requires a dedicated control line due to the low anharmonicity of transmon qubits, which prevents effective multiplexing for control signals. For flux-tunable qubit architectures, an additional flux line is required per qubit to enable frequency tuning and enable the execution of some single-qubit and two-qubit gates.

In the case of the remaining technologies, all the drive and readout wiring between the controlling unit and the qubit chip is removed, facilitating a scaling up of the number of qubits. Flux lines might still be needed, as is the case for wireless technologies, since these cannot be used in conjunction with the DC signals required to tune qubits to their optimal frequency. Nevertheless, using polarisation or modulation schemes with the wireless signals, it is potentially possible to address more than one qubit with the same biasing conditions, reducing the overhead further.
SFQ logic controllers, in particular, are the only technology in this review that can be used on the mK stage of the cryostat, in close proximity to the qubit chip. This means that all the wiring between these two units can be completely removed, and simple capacitive and inductive connections, either via flip-chip modules or with both units fabricated on the same substrate, facilitate the interaction between them\cite{Leonard2019DigitalQubit,Liu2023SingleModule}.

Another critical metric is the power consumption required to operate a single qubit \cite{PRXQuantum.4.040319}, which encompasses not only the signals themselves but also the energy demands of the supporting electronic systems. Particularly for CMOS technologies, the values shown in Table \ref{tab:table1} indicate the total AC and DC power provided to an FPGA and RF module to generate a stream of control pulses at a frequency around 1 GHz \cite{8865458}. 
The power requirements can vary dramatically across different technologies, but their impact also depends significantly on the temperature stage at which the electronics are located. For instance, while room-temperature electronics dissipate substantial heat, their scalability is practically unconstrained by thermal considerations. In contrast, cryo-CMOS electronics operating at the 4 K stage must adhere to stringent cooling power limitations to ensure proper functionality, as the available cooling power diminishes with decreasing temperature. In the case of SFQ logic and because of the superconducting nature of these circuits, the power dissipated is given solely by the dynamical switching of the Josephson junctions and is proportional to their critical current and the frequency of operation \cite{McDermott2018Quantum-classicalLogic,Barbosa2024}. 
Optical links remove the need for coaxial wiring and attenuators along the control lines, instead replacing it with simple and efficient optical fibers. Here, the majority of energy losses occur during the conversion between the optical and RF domains, with the RF-to-optical conversion being particularly inefficient. For instance, using a projected performance of $V_{\pi} = 50$ mV, insertion loss of 2.5 dB per facet and waveguide loss of 0.05 dB/cm \cite{Shen24photonicSFQ}, the power dissipation is in the order of tens of \textmu W per qubit. In comparison, optical-to-RF photodiode conversion achieves power dissipation in the nW range under similar conditions. Both scenarios assume a 1\% duty cycle, where the laser is active only during qubit addressing.
Wireless technology also removes the need of attenuators in the drive lines, allowing the usage of much lower power microwave circuitry, which are able to generate and transmit single photon power signals directly from on-chip antennas.

\section*{Conclusion} \label{conclusion}
\begin{figure}
  \centering
  \includegraphics[scale=0.55, trim=0 0 0 2cm]{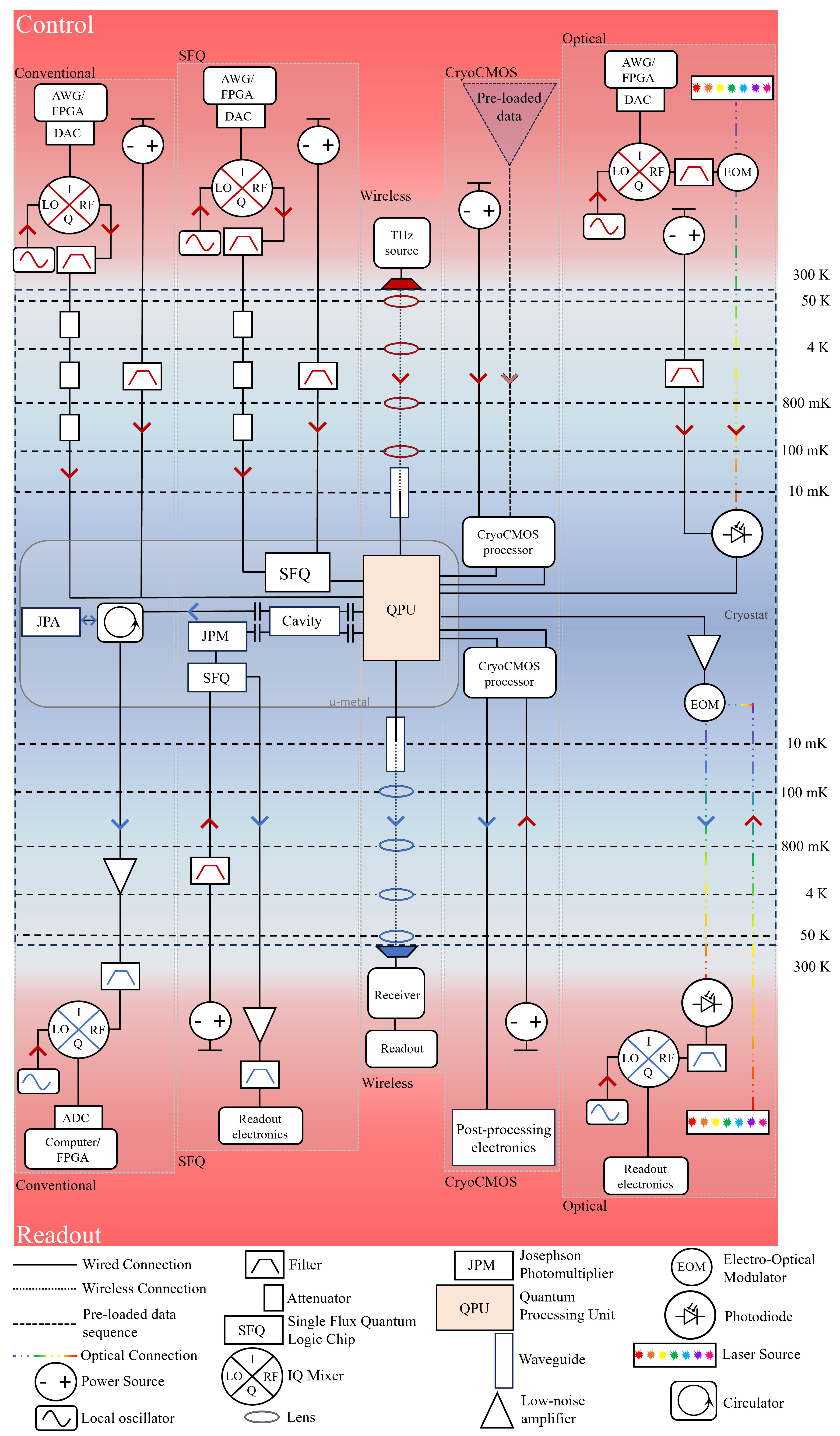}
    \caption{Simplified wiring schematics of leading control (top) and readout (bottom) architectures for superconducting quantum processing units (QPUs). From left to right: conventional, single flux quantum (SFQ), wireless, CryoCMOS, and optical interfaces.}
    \label{fig:Full Schematic}
\end{figure}

We have reviewed the current state of the art in terms of the classical interfaces required to control cryogenic quantum systems. A larger schematic, giving a simplified overview of each of these potential interfaces is shown in Figure \ref{fig:Full Schematic} Given the unique strengths and limitations of each technology, it is likely that future large-scale quantum devices will leverage a combination of these technologies, strategically applied at different stages of the setup to maximize their respective advantages. Quantum devices have huge potential across a number of fields and could represent a paradigm shift in many areas of technology. However, control systems for these devices are currently at a very early stage of development, and progress needs to be made to keep up with the rapid advances in areas such as qubit fabrication and quantum algorithms or software.

\clearpage
\newgeometry{margin=3cm, top=1cm, bottom=1cm}
\thispagestyle{empty}
\clearpage
\restoregeometry

\backmatter


\bmhead{Acknowledgments}
This work was supported by the EPSRC QCS Hub (PRF-02-A-03), UKRI Innovate UK Cryo-CMOS to enable scalable quantum computers project (grant no. 10006017), UKRI Innovate UK Altnaharra: Cryoelectronics for Quantum Circuits project (grant no. 10006186) and UKRI EPSRC Empowering Practical Interfacing of Quantum Computing (EPIQC) (grant no. EP/W032627/1)

\section*{Declarations}
The authors declare no conflicts of interest in this work.

\begin{appendices}





\end{appendices}

\bibliography{QuantClassRefs_1}


\begin{thebibliography}{114}
\ifx \bisbn   \undefined \def \bisbn  #1{ISBN #1}\fi
\ifx \binits  \undefined \def \binits#1{#1}\fi
\ifx \bauthor  \undefined \def \bauthor#1{#1}\fi
\ifx \batitle  \undefined \def \batitle#1{#1}\fi
\ifx \bjtitle  \undefined \def \bjtitle#1{#1}\fi
\ifx \bvolume  \undefined \def \bvolume#1{\textbf{#1}}\fi
\ifx \byear  \undefined \def \byear#1{#1}\fi
\ifx \bissue  \undefined \def \bissue#1{#1}\fi
\ifx \bfpage  \undefined \def \bfpage#1{#1}\fi
\ifx \blpage  \undefined \def \blpage #1{#1}\fi
\ifx \burl  \undefined \def \burl#1{\textsf{#1}}\fi
\ifx \doiurl  \undefined \def \doiurl#1{\url{https://doi.org/#1}}\fi
\ifx \betal  \undefined \def \betal{\textit{et al.}}\fi
\ifx \binstitute  \undefined \def \binstitute#1{#1}\fi
\ifx \binstitutionaled  \undefined \def \binstitutionaled#1{#1}\fi
\ifx \bctitle  \undefined \def \bctitle#1{#1}\fi
\ifx \beditor  \undefined \def \beditor#1{#1}\fi
\ifx \bpublisher  \undefined \def \bpublisher#1{#1}\fi
\ifx \bbtitle  \undefined \def \bbtitle#1{#1}\fi
\ifx \bedition  \undefined \def \bedition#1{#1}\fi
\ifx \bseriesno  \undefined \def \bseriesno#1{#1}\fi
\ifx \blocation  \undefined \def \blocation#1{#1}\fi
\ifx \bsertitle  \undefined \def \bsertitle#1{#1}\fi
\ifx \bsnm \undefined \def \bsnm#1{#1}\fi
\ifx \bsuffix \undefined \def \bsuffix#1{#1}\fi
\ifx \bparticle \undefined \def \bparticle#1{#1}\fi
\ifx \barticle \undefined \def \barticle#1{#1}\fi
\bibcommenthead
\ifx \bconfdate \undefined \def \bconfdate #1{#1}\fi
\ifx \botherref \undefined \def \botherref #1{#1}\fi
\ifx \url \undefined \def \url#1{\textsf{#1}}\fi
\ifx \bchapter \undefined \def \bchapter#1{#1}\fi
\ifx \bbook \undefined \def \bbook#1{#1}\fi
\ifx \bcomment \undefined \def \bcomment#1{#1}\fi
\ifx \oauthor \undefined \def \oauthor#1{#1}\fi
\ifx \citeauthoryear \undefined \def \citeauthoryear#1{#1}\fi
\ifx \endbibitem  \undefined \def \endbibitem {}\fi
\ifx \bconflocation  \undefined \def \bconflocation#1{#1}\fi
\ifx \arxivurl  \undefined \def \arxivurl#1{\textsf{#1}}\fi
\csname PreBibitemsHook\endcsname

\bibitem[\protect\citeauthoryear{Steane}{1998}]{Steane1998}
\begin{barticle}
\bauthor{\bsnm{Steane}, \binits{A.}}:
\batitle{Quantum computing}.
\bjtitle{Reports on Progress in Physics}
\bvolume{61},
\bfpage{117}
(\byear{1998})
\doiurl{10.1088/0034-4885/61/2/002}
\end{barticle}
\endbibitem

\bibitem[\protect\citeauthoryear{Spentzouris}{2020}]{Spentzouris2020}
\begin{botherref}
\oauthor{\bsnm{Spentzouris}, \binits{P.}}:
Quantum Computing: Advancing Fundamental Physics.
Springer
(2020).
\doiurl{10.1007/s41781-020-00043-x}
\end{botherref}
\endbibitem

\bibitem[\protect\citeauthoryear{Bonde et~al.}{2024}]{Bonde2024}
\begin{bbook}
\bauthor{\bsnm{Bonde}, \binits{B.}},
\bauthor{\bsnm{Patil}, \binits{P.}},
\bauthor{\bsnm{Choubey}, \binits{B.}}:
In: \beditor{\bsnm{Heifetz}, \binits{A.}} (ed.)
\bbtitle{The Future of Drug Development with Quantum Computing},
pp. \bfpage{153}--\blpage{179}.
\bpublisher{Springer}, \blocation{???}
(\byear{2024}).
\doiurl{10.1007/978-1-0716-3449-3_7}
\end{bbook}
\endbibitem

\bibitem[\protect\citeauthoryear{MacQuarrie et~al.}{2020}]{MacQuarrie2020}
\begin{barticle}
\bauthor{\bsnm{MacQuarrie}, \binits{E.R.}},
\bauthor{\bsnm{Simon}, \binits{C.}},
\bauthor{\bsnm{Simmons}, \binits{S.}},
\bauthor{\bsnm{Maine}, \binits{E.}}:
\batitle{The emerging commercial landscape of quantum computing}.
\bjtitle{Nature Reviews Physics}
\bvolume{2},
\bfpage{596}--\blpage{598}
(\byear{2020})
\doiurl{10.1038/s42254-020-00247-5}
\end{barticle}
\endbibitem

\bibitem[\protect\citeauthoryear{Fedorov et~al.}{2022}]{Fedorov2022}
\begin{botherref}
\oauthor{\bsnm{Fedorov}, \binits{A.K.}},
\oauthor{\bsnm{Gisin}, \binits{N.}},
\oauthor{\bsnm{Beloussov}, \binits{S.M.}},
\oauthor{\bsnm{Lvovsky}, \binits{A.I.}}:
Quantum computing at the quantum advantage threshold: a down-to-business review
(2022)
\doiurl{10.48550/arXiv.2203.17181}
\end{botherref}
\endbibitem

\bibitem[\protect\citeauthoryear{Almudever et~al.}{2017}]{Almudever2017}
\begin{bchapter}
\bauthor{\bsnm{Almudever}, \binits{C.G.}},
\bauthor{\bsnm{Lao}, \binits{L.}},
\bauthor{\bsnm{Fu}, \binits{X.}},
\bauthor{\bsnm{Khammassi}, \binits{N.}},
\bauthor{\bsnm{Ashraf}, \binits{I.}},
\bauthor{\bsnm{Iorga}, \binits{D.}},
\bauthor{\bsnm{Varsamopoulos}, \binits{S.}},
\bauthor{\bsnm{Eichler}, \binits{C.}},
\bauthor{\bsnm{Wallraff}, \binits{A.}},
\bauthor{\bsnm{Geck}, \binits{L.}},
\bauthor{\bsnm{Kruth}, \binits{A.}},
\bauthor{\bsnm{Knoch}, \binits{J.}},
\bauthor{\bsnm{Bluhm}, \binits{H.}},
\bauthor{\bsnm{Bertels}, \binits{K.}}:
\bctitle{The engineering challenges in quantum computing},
pp. \bfpage{836}--\blpage{845}
(\byear{2017}).
\doiurl{10.23919/DATE.2017.7927104}
\end{bchapter}
\endbibitem

\bibitem[\protect\citeauthoryear{Preskill}{2018}]{Preskill2018}
\begin{barticle}
\bauthor{\bsnm{Preskill}, \binits{J.}}:
\batitle{Quantum {C}omputing in the {NISQ} era and beyond}.
\bjtitle{{Quantum}}
\bvolume{2},
\bfpage{79}
(\byear{2018})
\doiurl{10.22331/q-2018-08-06-79}
\end{barticle}
\endbibitem

\bibitem[\protect\citeauthoryear{{IBM}}{2023}]{IBM2023IBM:Qubits}
\begin{botherref}
\oauthor{\bsnm{{IBM}}}:
{IBM: Charting the course to 100,000 qubits}.
\url{https://research.ibm.com/blog/100k-qubit-supercomputer}
(2023)
\end{botherref}
\endbibitem

\bibitem[\protect\citeauthoryear{{UK Department for Science, Innovation, and Technology}}{2023}]{UKDepartmentforScience2023UKStrategy}
\begin{botherref}
\oauthor{\bsnm{{UK Department for Science, Innovation, and Technology}}}:
{UK Government: National Quantum Strategy}.
\url{https://www.gov.uk/government/publications/national-quantum-strategy}
(2023)
\end{botherref}
\endbibitem

\bibitem[\protect\citeauthoryear{{European Commission}}{2022}]{EuropeanCommission2022EU:Flagship}
\begin{botherref}
\oauthor{\bsnm{{European Commission}}}:
{EU: Quantum Technologies Flagship}.
\url{https://digital-strategy.ec.europa.eu/en/policies/quantum-technologies-flagship}
(2022)
\end{botherref}
\endbibitem

\bibitem[\protect\citeauthoryear{Gill et~al.}{2022}]{Gill2022}
\begin{barticle}
\bauthor{\bsnm{Gill}, \binits{S.S.}},
\bauthor{\bsnm{Kumar}, \binits{A.}},
\bauthor{\bsnm{Singh}, \binits{H.}},
\bauthor{\bsnm{Singh}, \binits{M.}},
\bauthor{\bsnm{Kaur}, \binits{K.}},
\bauthor{\bsnm{Usman}, \binits{M.}},
\bauthor{\bsnm{Buyya}, \binits{R.}}:
\batitle{Quantum computing: A taxonomy, systematic review and future directions}.
\bjtitle{Software: Practice and Experience}
\bvolume{52},
\bfpage{66}--\blpage{114}
(\byear{2022})
\doiurl{10.1002/spe.3039}
\end{barticle}
\endbibitem

\bibitem[\protect\citeauthoryear{{National Academies of Sciences, Engineering, and Medicine}}{2019}]{horowitz2019quantum}
\begin{bbook}
\bauthor{\bsnm{{National Academies of Sciences, Engineering, and Medicine}}}:
\bbtitle{Quantum Computing: Progress and Prospects}.
\bpublisher{The National Academies Press},
\blocation{Washington, DC}
(\byear{2019}).
\doiurl{10.17226/25196} .
\burl{https://nap.nationalacademies.org/catalog/25196/quantum-computing-progress-and-prospects}
\end{bbook}
\endbibitem

\bibitem[\protect\citeauthoryear{Siddiqi}{2021}]{Siddiqi2021}
\begin{barticle}
\bauthor{\bsnm{Siddiqi}, \binits{I.}}:
\batitle{Engineering high-coherence superconducting qubits}.
\bjtitle{Nature Reviews Materials}
\bvolume{6},
\bfpage{875}--\blpage{891}
(\byear{2021})
\doiurl{10.1038/s41578-021-00370-4}
\end{barticle}
\endbibitem

\bibitem[\protect\citeauthoryear{de~Leon et~al.}{2021}]{deLeon2021}
\begin{barticle}
\bauthor{\bsnm{Leon}, \binits{N.P.}},
\bauthor{\bsnm{Itoh}, \binits{K.M.}},
\bauthor{\bsnm{Kim}, \binits{D.}},
\bauthor{\bsnm{Mehta}, \binits{K.K.}},
\bauthor{\bsnm{Northup}, \binits{T.E.}},
\bauthor{\bsnm{Paik}, \binits{H.}},
\bauthor{\bsnm{Palmer}, \binits{B.S.}},
\bauthor{\bsnm{Samarth}, \binits{N.}},
\bauthor{\bsnm{Sangtawesin}, \binits{S.}},
\bauthor{\bsnm{Steuerman}, \binits{D.W.}}:
\batitle{Materials challenges and opportunities for quantum computing hardware}.
\bjtitle{Science}
\bvolume{372}(\bissue{6539}),
\bfpage{2823}
(\byear{2021})
\doiurl{10.1126/science.abb2823}
{\href{https://arxiv.org/abs/https://www.science.org/doi/pdf/10.1126/science.abb2823}{{https://www.science.org/doi/pdf/10.1126/science.abb2823}}}
\end{barticle}
\endbibitem

\bibitem[\protect\citeauthoryear{Campbell}{2024}]{Campbell2024}
\begin{barticle}
\bauthor{\bsnm{Campbell}, \binits{E.}}:
\batitle{A series of fast-paced advances in quantum error correction}.
\bjtitle{Nature Reviews Physics}
\bvolume{6},
\bfpage{160}--\blpage{161}
(\byear{2024})
\doiurl{10.1038/s42254-024-00706-3}
\end{barticle}
\endbibitem

\bibitem[\protect\citeauthoryear{Bharti et~al.}{2022}]{Bharti2022}
\begin{barticle}
\bauthor{\bsnm{Bharti}, \binits{K.}},
\bauthor{\bsnm{Cervera-Lierta}, \binits{A.}},
\bauthor{\bsnm{Kyaw}, \binits{T.H.}},
\bauthor{\bsnm{Haug}, \binits{T.}},
\bauthor{\bsnm{Alperin-Lea}, \binits{S.}},
\bauthor{\bsnm{Anand}, \binits{A.}},
\bauthor{\bsnm{Degroote}, \binits{M.}},
\bauthor{\bsnm{Heimonen}, \binits{H.}},
\bauthor{\bsnm{Kottmann}, \binits{J.S.}},
\bauthor{\bsnm{Menke}, \binits{T.}},
\bauthor{\bsnm{Mok}, \binits{W.-K.}},
\bauthor{\bsnm{Sim}, \binits{S.}},
\bauthor{\bsnm{Kwek}, \binits{L.-C.}},
\bauthor{\bsnm{Aspuru-Guzik}, \binits{A.}}:
\batitle{Noisy intermediate-scale quantum algorithms}.
\bjtitle{Reviews of Modern Physics}
\bvolume{94},
\bfpage{15004}
(\byear{2022})
\doiurl{10.1103/RevModPhys.94.015004}
\end{barticle}
\endbibitem

\bibitem[\protect\citeauthoryear{Shammah et~al.}{2024}]{Shammah2024}
\begin{barticle}
\bauthor{\bsnm{Shammah}, \binits{N.}},
\bauthor{\bsnm{Saha~Roy}, \binits{A.}},
\bauthor{\bsnm{Almudever}, \binits{C.G.}},
\bauthor{\bsnm{Bourdeauducq}, \binits{S.}},
\bauthor{\bsnm{Butko}, \binits{A.}},
\bauthor{\bsnm{Cancelo}, \binits{G.}},
\bauthor{\bsnm{Clark}, \binits{S.M.}},
\bauthor{\bsnm{Heinsoo}, \binits{J.}},
\bauthor{\bsnm{Henriet}, \binits{L.}},
\bauthor{\bsnm{Huang}, \binits{G.}},
\bauthor{\bsnm{Jurczak}, \binits{C.}},
\bauthor{\bsnm{Kotilahti}, \binits{J.}},
\bauthor{\bsnm{Landra}, \binits{A.}},
\bauthor{\bsnm{LaRose}, \binits{R.}},
\bauthor{\bsnm{Mari}, \binits{A.}},
\bauthor{\bsnm{Nowrouzi}, \binits{K.}},
\bauthor{\bsnm{Ockeloen-Korppi}, \binits{C.}},
\bauthor{\bsnm{Prawiroatmodjo}, \binits{G.}},
\bauthor{\bsnm{Siddiqi}, \binits{I.}},
\bauthor{\bsnm{Zeng}, \binits{W.J.}}:
\batitle{{Open hardware solutions in quantum technology}}.
\bjtitle{APL Quantum}
\bvolume{1}(\bissue{1}),
\bfpage{011501}
(\byear{2024})
\doiurl{10.1063/5.0180987}
{\href{https://arxiv.org/abs/https://pubs.aip.org/aip/apq/article-pdf/doi/10.1063/5.0180987/19841793/011501\_1\_5.0180987.pdf}{{https://pubs.aip.org/aip/apq/article-pdf/doi/10.1063/5.0180987/19841793/011501\_1\_5.0180987.pdf}}}
\end{barticle}
\endbibitem

\bibitem[\protect\citeauthoryear{Cerezo et~al.}{2021}]{Cerezo2021}
\begin{barticle}
\bauthor{\bsnm{Cerezo}, \binits{M.}},
\bauthor{\bsnm{Arrasmith}, \binits{A.}},
\bauthor{\bsnm{Babbush}, \binits{R.}},
\bauthor{\bsnm{Benjamin}, \binits{S.C.}},
\bauthor{\bsnm{Endo}, \binits{S.}},
\bauthor{\bsnm{Fujii}, \binits{K.}},
\bauthor{\bsnm{McClean}, \binits{J.R.}},
\bauthor{\bsnm{Mitarai}, \binits{K.}},
\bauthor{\bsnm{Yuan}, \binits{X.}},
\bauthor{\bsnm{Cincio}, \binits{L.}},
\bauthor{\bsnm{Coles}, \binits{P.J.}}:
\batitle{Variational quantum algorithms}.
\bjtitle{Nature Reviews Physics}
\bvolume{3},
\bfpage{625}--\blpage{644}
(\byear{2021})
\doiurl{10.1038/s42254-021-00348-9}
\end{barticle}
\endbibitem

\bibitem[\protect\citeauthoryear{Ezratty}{2023}]{Ezratty2023}
\begin{barticle}
\bauthor{\bsnm{Ezratty}, \binits{O.}}:
\batitle{Perspective on superconducting qubit quantum computing}.
\bjtitle{The European Physical Journal A}
\bvolume{59},
\bfpage{94}
(\byear{2023})
\doiurl{10.1140/epja/s10050-023-01006-7}
\end{barticle}
\endbibitem

\bibitem[\protect\citeauthoryear{Serniak et~al.}{2018}]{hotqp2018}
\begin{barticle}
\bauthor{\bsnm{Serniak}, \binits{K.}},
\bauthor{\bsnm{Hays}, \binits{M.}},
\bauthor{\bsnm{Lange}, \binits{G.}},
\bauthor{\bsnm{Diamond}, \binits{S.}},
\bauthor{\bsnm{Shankar}, \binits{S.}},
\bauthor{\bsnm{Burkhart}, \binits{L.D.}},
\bauthor{\bsnm{Frunzio}, \binits{L.}},
\bauthor{\bsnm{Houzet}, \binits{M.}},
\bauthor{\bsnm{Devoret}, \binits{M.H.}}:
\batitle{Hot nonequilibrium quasiparticles in transmon qubits}.
\bjtitle{Phys. Rev. Lett.}
\bvolume{121},
\bfpage{157701}
(\byear{2018})
\doiurl{10.1103/PhysRevLett.121.157701}
\end{barticle}
\endbibitem

\bibitem[\protect\citeauthoryear{Krinner et~al.}{2019}]{Krinner2019}
\begin{barticle}
\bauthor{\bsnm{Krinner}, \binits{S.}},
\bauthor{\bsnm{Storz}, \binits{S.}},
\bauthor{\bsnm{Kurpiers}, \binits{P.}},
\bauthor{\bsnm{Magnard}, \binits{P.}},
\bauthor{\bsnm{Heinsoo}, \binits{J.}},
\bauthor{\bsnm{Keller}, \binits{R.}},
\bauthor{\bsnm{Lütolf}, \binits{J.}},
\bauthor{\bsnm{Eichler}, \binits{C.}},
\bauthor{\bsnm{Wallraff}, \binits{A.}}:
\batitle{Engineering cryogenic setups for 100-qubit scale superconducting circuit systems}.
\bjtitle{EPJ Quantum Technology}
\bvolume{6}(\bissue{1}),
\bfpage{2}
(\byear{2019})
\doiurl{10.1140/epjqt/s40507-019-0072-0}
\end{barticle}
\endbibitem

\bibitem[\protect\citeauthoryear{Clarke and Wilhelm}{2008}]{Clarke2008SuperconductingBits}
\begin{barticle}
\bauthor{\bsnm{Clarke}, \binits{J.}},
\bauthor{\bsnm{Wilhelm}, \binits{F.K.}}:
\batitle{{Superconducting quantum bits}}.
\bjtitle{Nature}
\bvolume{453}(\bissue{7198}),
\bfpage{1031}--\blpage{1042}
(\byear{2008})
\doiurl{10.1038/nature07128}
\end{barticle}
\endbibitem

\bibitem[\protect\citeauthoryear{Wendin}{2017}]{Wendin2017}
\begin{barticle}
\bauthor{\bsnm{Wendin}, \binits{G.}}:
\batitle{Quantum information processing with superconducting circuits: a review}.
\bjtitle{Reports on Progress in Physics}
\bvolume{80},
\bfpage{106001}
(\byear{2017})
\doiurl{10.1088/1361-6633/aa7e1a}
\end{barticle}
\endbibitem

\bibitem[\protect\citeauthoryear{Verjauw et~al.}{2022}]{Verjauw2022}
\begin{barticle}
\bauthor{\bsnm{Verjauw}, \binits{J.}},
\bauthor{\bsnm{Acharya}, \binits{R.}},
\bauthor{\bsnm{Damme}, \binits{J.V.}},
\bauthor{\bsnm{Ivanov}, \binits{T.}},
\bauthor{\bsnm{Lozano}, \binits{D.P.}},
\bauthor{\bsnm{Mohiyaddin}, \binits{F.A.}},
\bauthor{\bsnm{Wan}, \binits{D.}},
\bauthor{\bsnm{Jussot}, \binits{J.}},
\bauthor{\bsnm{Vadiraj}, \binits{A.M.}},
\bauthor{\bsnm{Mongillo}, \binits{M.}},
\bauthor{\bsnm{Heyns}, \binits{M.}},
\bauthor{\bsnm{Radu}, \binits{I.}},
\bauthor{\bsnm{Govoreanu}, \binits{B.}},
\bauthor{\bsnm{Potočnik}, \binits{A.}}:
\batitle{Path toward manufacturable superconducting qubits with relaxation times exceeding 0.1 ms}.
\bjtitle{npj Quantum Information}
\bvolume{8},
\bfpage{93}
(\byear{2022})
\doiurl{10.1038/s41534-022-00600-9}
\end{barticle}
\endbibitem

\bibitem[\protect\citeauthoryear{Van~Damme et~al.}{2024}]{VanDamme2024}
\begin{barticle}
\bauthor{\bsnm{Van~Damme}, \binits{J.}},
\bauthor{\bsnm{Massar}, \binits{S.}},
\bauthor{\bsnm{Acharya}, \binits{R.}},
\bauthor{\bsnm{Ivanov}, \binits{T.}},
\bauthor{\bsnm{Perez~Lozano}, \binits{D.}},
\bauthor{\bsnm{Canvel}, \binits{Y.}},
\bauthor{\bsnm{Demarets}, \binits{M.}},
\bauthor{\bsnm{Vangoidsenhoven}, \binits{D.}},
\bauthor{\bsnm{Hermans}, \binits{Y.}},
\bauthor{\bsnm{Lai}, \binits{J.G.}},
\bauthor{\bsnm{Vadiraj}, \binits{A.M.}},
\bauthor{\bsnm{Mongillo}, \binits{M.}},
\bauthor{\bsnm{Wan}, \binits{D.}},
\bauthor{\bsnm{De~Boeck}, \binits{J.}},
\bauthor{\bsnm{Potočnik}, \binits{A.}},
\bauthor{\bsnm{De~Greve}, \binits{K.}}:
\batitle{Advanced cmos manufacturing of superconducting qubits on 300 mm wafers}.
\bjtitle{Nature}
\bvolume{634}(\bissue{8032}),
\bfpage{74}--\blpage{79}
(\byear{2024})
\doiurl{10.1038/s41586-024-07941-9}
\end{barticle}
\endbibitem

\bibitem[\protect\citeauthoryear{Kjaergaard et~al.}{2020}]{Kjaergaard2020SuperconductingPlay}
\begin{botherref}
\oauthor{\bsnm{Kjaergaard}, \binits{M.}},
\oauthor{\bsnm{Schwartz}, \binits{M.E.}},
\oauthor{\bsnm{Braum{\"{u}}ller}, \binits{J.}},
\oauthor{\bsnm{Krantz}, \binits{P.}},
\oauthor{\bsnm{I-J~Wang}, \binits{J.}},
\oauthor{\bsnm{Gustavsson}, \binits{S.}},
\oauthor{\bsnm{Oliver}, \binits{W.D.}}:
{Superconducting Qubits: Current State of Play}
(2020)
\doiurl{10.1146/annurev-conmatphys}
\end{botherref}
\endbibitem

\bibitem[\protect\citeauthoryear{Oliver and Welander}{2013}]{Oliver2013}
\begin{barticle}
\bauthor{\bsnm{Oliver}, \binits{W.D.}},
\bauthor{\bsnm{Welander}, \binits{P.B.}}:
\batitle{Materials in superconducting quantum bits}.
\bjtitle{MRS Bulletin}
\bvolume{38},
\bfpage{816}--\blpage{825}
(\byear{2013})
\doiurl{10.1557/mrs.2013.229}
\end{barticle}
\endbibitem

\bibitem[\protect\citeauthoryear{Krantz et~al.}{2019}]{Krantz2019AQubits}
\begin{botherref}
\oauthor{\bsnm{Krantz}, \binits{P.}},
\oauthor{\bsnm{Kjaergaard}, \binits{M.}},
\oauthor{\bsnm{Yan}, \binits{F.}},
\oauthor{\bsnm{Orlando}, \binits{T.P.}},
\oauthor{\bsnm{Gustavsson}, \binits{S.}},
\oauthor{\bsnm{Oliver}, \binits{W.D.}}:
{A quantum engineer's guide to superconducting qubits}.
Applied Physics Reviews
\textbf{6}(2)
(2019)
\doiurl{10.1063/1.5089550}
\end{botherref}
\endbibitem

\bibitem[\protect\citeauthoryear{Philips et~al.}{2022}]{Philips2022UniversalSilicon}
\begin{barticle}
\bauthor{\bsnm{Philips}, \binits{S.G.J.}},
\bauthor{\bsnm{M{\c{a}}dzik}, \binits{M.T.}},
\bauthor{\bsnm{Amitonov}, \binits{S.V.}},
\bauthor{\bsnm{Snoo}, \binits{S.L.}},
\bauthor{\bsnm{Russ}, \binits{M.}},
\bauthor{\bsnm{Kalhor}, \binits{N.}},
\bauthor{\bsnm{Volk}, \binits{C.}},
\bauthor{\bsnm{Lawrie}, \binits{W.I.L.}},
\bauthor{\bsnm{Brousse}, \binits{D.}},
\bauthor{\bsnm{Tryputen}, \binits{L.}},
\bauthor{\bsnm{Wuetz}, \binits{B.P.}},
\bauthor{\bsnm{Sammak}, \binits{A.}},
\bauthor{\bsnm{Veldhorst}, \binits{M.}},
\bauthor{\bsnm{Scappucci}, \binits{G.}},
\bauthor{\bsnm{Vandersypen}, \binits{L.M.K.}}:
\batitle{{Universal control of a six-qubit quantum processor in silicon}}.
\bjtitle{Nature}
\bvolume{609}(\bissue{7929}),
\bfpage{919}--\blpage{924}
(\byear{2022})
\doiurl{10.1038/s41586-022-05117-x}
\end{barticle}
\endbibitem

\bibitem[\protect\citeauthoryear{Burkard et~al.}{2023}]{Burkard2023SemiconductorQubits}
\begin{botherref}
\oauthor{\bsnm{Burkard}, \binits{G.}},
\oauthor{\bsnm{Ladd}, \binits{T.D.}},
\oauthor{\bsnm{Pan}, \binits{A.}},
\oauthor{\bsnm{Nichol}, \binits{J.M.}},
\oauthor{\bsnm{Petta}, \binits{J.R.}}:
{Semiconductor spin qubits}.
Reviews of Modern Physics
\textbf{95}(2)
(2023)
\doiurl{10.1103/RevModPhys.95.025003}
\end{botherref}
\endbibitem

\bibitem[\protect\citeauthoryear{Bruzewicz et~al.}{2019}]{Bruzewicz2019Trapped-ionChallenges}
\begin{botherref}
\oauthor{\bsnm{Bruzewicz}, \binits{C.D.}},
\oauthor{\bsnm{Chiaverini}, \binits{J.}},
\oauthor{\bsnm{McConnell}, \binits{R.}},
\oauthor{\bsnm{Sage}, \binits{J.M.}}:
{Trapped-ion quantum computing: Progress and challenges}.
Applied Physics Reviews
\textbf{6}(2)
(2019)
\doiurl{10.1063/1.5088164}
\end{botherref}
\endbibitem

\bibitem[\protect\citeauthoryear{Slussarenko and Pryde}{2019}]{Slussarenko2019PhotonicReview}
\begin{botherref}
\oauthor{\bsnm{Slussarenko}, \binits{S.}},
\oauthor{\bsnm{Pryde}, \binits{G.J.}}:
{Photonic quantum information processing: A concise review}.
American Institute of Physics Inc.
(2019).
\doiurl{10.1063/1.5115814}
\end{botherref}
\endbibitem

\bibitem[\protect\citeauthoryear{Ramakrishnan et~al.}{2023}]{Ramakrishnan2023IntegratedReview}
\begin{barticle}
\bauthor{\bsnm{Ramakrishnan}, \binits{R.K.}},
\bauthor{\bsnm{Ravichandran}, \binits{A.B.}},
\bauthor{\bsnm{Mishra}, \binits{A.}},
\bauthor{\bsnm{Kaushalram}, \binits{A.}},
\bauthor{\bsnm{Hegde}, \binits{G.}},
\bauthor{\bsnm{Talabattula}, \binits{S.}},
\bauthor{\bsnm{Rohde}, \binits{P.P.}}:
\batitle{{Integrated photonic platforms for quantum technology: a review}}.
\bjtitle{ISSS Journal of Micro and Smart Systems}
\bvolume{12}(\bissue{2}),
\bfpage{83}--\blpage{104}
(\byear{2023})
\doiurl{10.1007/s41683-023-00115-1}
\end{barticle}
\endbibitem

\bibitem[\protect\citeauthoryear{Brandl et~al.}{2016}]{Brandl2016}
\begin{barticle}
\bauthor{\bsnm{Brandl}, \binits{M.F.}},
\bauthor{\bsnm{Mourik}, \binits{M.W.}},
\bauthor{\bsnm{Postler}, \binits{L.}},
\bauthor{\bsnm{Nolf}, \binits{A.}},
\bauthor{\bsnm{Lakhmanskiy}, \binits{K.}},
\bauthor{\bsnm{Paiva}, \binits{R.R.}},
\bauthor{\bsnm{Möller}, \binits{S.}},
\bauthor{\bsnm{Daniilidis}, \binits{N.}},
\bauthor{\bsnm{Häffner}, \binits{H.}},
\bauthor{\bsnm{Kaushal}, \binits{V.}},
\bauthor{\bsnm{Ruster}, \binits{T.}},
\bauthor{\bsnm{Warschburger}, \binits{C.}},
\bauthor{\bsnm{Kaufmann}, \binits{H.}},
\bauthor{\bsnm{Poschinger}, \binits{U.G.}},
\bauthor{\bsnm{Schmidt-Kaler}, \binits{F.}},
\bauthor{\bsnm{Schindler}, \binits{P.}},
\bauthor{\bsnm{Monz}, \binits{T.}},
\bauthor{\bsnm{Blatt}, \binits{R.}}:
\batitle{Cryogenic setup for trapped ion quantum computing}.
\bjtitle{Review of Scientific Instruments}
\bvolume{87},
\bfpage{113103}
(\byear{2016})
\doiurl{10.1063/1.4966970}
\end{barticle}
\endbibitem

\bibitem[\protect\citeauthoryear{Pagano et~al.}{2019}]{Pagano2019}
\begin{barticle}
\bauthor{\bsnm{Pagano}, \binits{G.}},
\bauthor{\bsnm{Hess}, \binits{P.W.}},
\bauthor{\bsnm{Kaplan}, \binits{H.B.}},
\bauthor{\bsnm{Tan}, \binits{W.L.}},
\bauthor{\bsnm{Richerme}, \binits{P.}},
\bauthor{\bsnm{Becker}, \binits{P.}},
\bauthor{\bsnm{Kyprianidis}, \binits{A.}},
\bauthor{\bsnm{Zhang}, \binits{J.}},
\bauthor{\bsnm{Birckelbaw}, \binits{E.}},
\bauthor{\bsnm{Hernandez}, \binits{M.R.}},
\bauthor{\bsnm{Wu}, \binits{Y.}},
\bauthor{\bsnm{Monroe}, \binits{C.}}:
\batitle{Cryogenic trapped-ion system for large scale quantum simulation}.
\bjtitle{Quantum Science and Technology}
\bvolume{4},
\bfpage{014004}
(\byear{2019})
\doiurl{10.1088/2058-9565/aae0fe}
\end{barticle}
\endbibitem

\bibitem[\protect\citeauthoryear{Brown et~al.}{2021}]{Brown2021}
\begin{barticle}
\bauthor{\bsnm{Brown}, \binits{K.R.}},
\bauthor{\bsnm{Chiaverini}, \binits{J.}},
\bauthor{\bsnm{Sage}, \binits{J.M.}},
\bauthor{\bsnm{Häffner}, \binits{H.}}:
\batitle{Materials challenges for trapped-ion quantum computers}.
\bjtitle{Nature Reviews Materials}
\bvolume{6},
\bfpage{892}--\blpage{905}
(\byear{2021})
\doiurl{10.1038/s41578-021-00292-1}
\end{barticle}
\endbibitem

\bibitem[\protect\citeauthoryear{Todaro et~al.}{2021}]{Todaro2021}
\begin{barticle}
\bauthor{\bsnm{Todaro}, \binits{S.L.}},
\bauthor{\bsnm{Verma}, \binits{V.B.}},
\bauthor{\bsnm{McCormick}, \binits{K.C.}},
\bauthor{\bsnm{Allcock}, \binits{D.T.C.}},
\bauthor{\bsnm{Mirin}, \binits{R.P.}},
\bauthor{\bsnm{Wineland}, \binits{D.J.}},
\bauthor{\bsnm{Nam}, \binits{S.W.}},
\bauthor{\bsnm{Wilson}, \binits{A.C.}},
\bauthor{\bsnm{Leibfried}, \binits{D.}},
\bauthor{\bsnm{Slichter}, \binits{D.H.}}:
\batitle{State readout of a trapped ion qubit using a trap-integrated superconducting photon detector}.
\bjtitle{Phys. Rev. Lett.}
\bvolume{126},
\bfpage{010501}
(\byear{2021})
\doiurl{10.1103/PhysRevLett.126.010501}
\end{barticle}
\endbibitem

\bibitem[\protect\citeauthoryear{Gyger et~al.}{2021}]{Gyger2021}
\begin{barticle}
\bauthor{\bsnm{Gyger}, \binits{S.}},
\bauthor{\bsnm{Zichi}, \binits{J.}},
\bauthor{\bsnm{Schweickert}, \binits{L.}},
\bauthor{\bsnm{Elshaari}, \binits{A.W.}},
\bauthor{\bsnm{Steinhauer}, \binits{S.}},
\bauthor{\bsnm{Silva}, \binits{S.F.C.}},
\bauthor{\bsnm{Rastelli}, \binits{A.}},
\bauthor{\bsnm{Zwiller}, \binits{V.}},
\bauthor{\bsnm{Jöns}, \binits{K.D.}},
\bauthor{\bsnm{Errando-Herranz}, \binits{C.}}:
\batitle{Reconfigurable photonics with on-chip single-photon detectors}.
\bjtitle{Nature Communications}
\bvolume{12},
\bfpage{1408}
(\byear{2021})
\doiurl{10.1038/s41467-021-21624-3}
\end{barticle}
\endbibitem

\bibitem[\protect\citeauthoryear{Esmaeil~Zadeh et~al.}{2021}]{Zadeh2021}
\begin{barticle}
\bauthor{\bsnm{Esmaeil~Zadeh}, \binits{I.}},
\bauthor{\bsnm{Chang}, \binits{J.}},
\bauthor{\bsnm{Los}, \binits{J.W.N.}},
\bauthor{\bsnm{Gyger}, \binits{S.}},
\bauthor{\bsnm{Elshaari}, \binits{A.W.}},
\bauthor{\bsnm{Steinhauer}, \binits{S.}},
\bauthor{\bsnm{Dorenbos}, \binits{S.N.}},
\bauthor{\bsnm{Zwiller}, \binits{V.}}:
\batitle{{Superconducting nanowire single-photon detectors: A perspective on evolution, state-of-the-art, future developments, and applications}}.
\bjtitle{Applied Physics Letters}
\bvolume{118}(\bissue{19}),
\bfpage{190502}
(\byear{2021})
\doiurl{10.1063/5.0045990}
{\href{https://arxiv.org/abs/https://pubs.aip.org/aip/apl/article-pdf/doi/10.1063/5.0045990/14547612/190502\_1\_online.pdf}{{https://pubs.aip.org/aip/apl/article-pdf/doi/10.1063/5.0045990/14547612/190502\_1\_online.pdf}}}
\end{barticle}
\endbibitem

\bibitem[\protect\citeauthoryear{Byun et~al.}{2021}]{Byun2021}
\begin{barticle}
\bauthor{\bsnm{Byun}, \binits{I.}},
\bauthor{\bsnm{Min}, \binits{D.}},
\bauthor{\bsnm{Lee}, \binits{G.}},
\bauthor{\bsnm{Na}, \binits{S.}},
\bauthor{\bsnm{Kim}, \binits{J.}}:
\batitle{A next-generation cryogenic processor architecture}.
\bjtitle{IEEE Micro}
\bvolume{41}(\bissue{3}),
\bfpage{80}--\blpage{86}
(\byear{2021})
\doiurl{10.1109/MM.2021.3070133}
\end{barticle}
\endbibitem

\bibitem[\protect\citeauthoryear{Holmes}{2021}]{CE_QIP_RM2022}
\begin{bchapter}
\bauthor{\bsnm{Holmes}, \binits{D.}}:
\bctitle{Cryogenic electronics and quantum information processing}.
In: \bbtitle{2021 IEEE International Roadmap for Devices and Systems Outbriefs},
pp. \bfpage{1}--\blpage{93}.
\bpublisher{IEEE Computer Society},
\blocation{Los Alamitos, CA, USA}
(\byear{2021}).
\doiurl{10.1109/IRDS54852.2021.00012} .
\burl{https://doi.ieeecomputersociety.org/10.1109/IRDS54852.2021.00012}
\end{bchapter}
\endbibitem

\bibitem[\protect\citeauthoryear{Motzoi et~al.}{2009}]{controlDRAG}
\begin{barticle}
\bauthor{\bsnm{Motzoi}, \binits{F.}},
\bauthor{\bsnm{Gambetta}, \binits{J.M.}},
\bauthor{\bsnm{Rebentrost}, \binits{P.}},
\bauthor{\bsnm{Wilhelm}, \binits{F.K.}}:
\batitle{Simple pulses for elimination of leakage in weakly nonlinear qubits}.
\bjtitle{Phys. Rev. Lett.}
\bvolume{103},
\bfpage{110501}
(\byear{2009})
\doiurl{10.1103/PhysRevLett.103.110501}
\end{barticle}
\endbibitem

\bibitem[\protect\citeauthoryear{Ezzell et~al.}{2023}]{controlDD}
\begin{barticle}
\bauthor{\bsnm{Ezzell}, \binits{N.}},
\bauthor{\bsnm{Pokharel}, \binits{B.}},
\bauthor{\bsnm{Tewala}, \binits{L.}},
\bauthor{\bsnm{Quiroz}, \binits{G.}},
\bauthor{\bsnm{Lidar}, \binits{D.A.}}:
\batitle{Dynamical decoupling for superconducting qubits: A performance survey}.
\bjtitle{Phys. Rev. Appl.}
\bvolume{20},
\bfpage{064027}
(\byear{2023})
\doiurl{10.1103/PhysRevApplied.20.064027}
\end{barticle}
\endbibitem

\bibitem[\protect\citeauthoryear{Pozar}{2011}]{Pozar2011MicrowaveEngineering}
\begin{bbook}
\bauthor{\bsnm{Pozar}, \binits{D.M.}}:
\bbtitle{{Microwave Engineering}},
\bedition{4th} edn.
\bpublisher{John Wiley {\&} Sons, Inc},
\blocation{Hoboken}
(\byear{2011})
\end{bbook}
\endbibitem

\bibitem[\protect\citeauthoryear{Caves}{1982}]{Caves1982QuantumAmplifiers}
\begin{barticle}
\bauthor{\bsnm{Caves}, \binits{C.M.}}:
\batitle{{Quantum limits on noise in linear amplifiers}}.
\bjtitle{Physical Review D}
\bvolume{26}(\bissue{8}),
\bfpage{1817}--\blpage{1839}
(\byear{1982})
\doiurl{10.1103/PhysRevD.26.1817}
\end{barticle}
\endbibitem

\bibitem[\protect\citeauthoryear{Leonhardt and Paul}{1993}]{Leonhardt1993RealisticDistributions}
\begin{botherref}
\oauthor{\bsnm{Leonhardt}, \binits{U.}},
\oauthor{\bsnm{Paul}, \binits{H.}}:
{Realistic optical homodyne measurements and quasiprobability distributions}.
Physical Review A
\textbf{48}(6)
(1993)
\end{botherref}
\endbibitem

\bibitem[\protect\citeauthoryear{Wiseman and Milburn}{2009}]{Wiseman2009QuantumControl}
\begin{bbook}
\bauthor{\bsnm{Wiseman}, \binits{H.M.}},
\bauthor{\bsnm{Milburn}, \binits{G.J.}}:
\bbtitle{{Quantum Measurement and Control}}.
\bpublisher{Cambridge University Press}, \blocation{???}
(\byear{2009}).
\doiurl{10.1017/CBO9780511813948} .
\burl{https://www.cambridge.org/core/product/identifier/9780511813948/type/book}
\end{bbook}
\endbibitem

\bibitem[\protect\citeauthoryear{{Entegra}}{2024}]{EntegraFPGA}
\begin{botherref}
\oauthor{\bsnm{{Entegra}}}:
{Entegra RFX-8440 - FPGA Zynq UltraScale+ RFSoC}
(2024).
\url{https://www.entegra.co.uk/ii-products/rfsoc-rfx-8440-data-acquisition-fpga-card/}
\end{botherref}
\endbibitem

\bibitem[\protect\citeauthoryear{Salath{\'{e}} et~al.}{2018}]{Salathe2018Low-LatencyCommunication}
\begin{botherref}
\oauthor{\bsnm{Salath{\'{e}}}, \binits{Y.}},
\oauthor{\bsnm{Kurpiers}, \binits{P.}},
\oauthor{\bsnm{Karg}, \binits{T.}},
\oauthor{\bsnm{Lang}, \binits{C.}},
\oauthor{\bsnm{Andersen}, \binits{C.K.}},
\oauthor{\bsnm{Akin}, \binits{A.}},
\oauthor{\bsnm{Krinner}, \binits{S.}},
\oauthor{\bsnm{Eichler}, \binits{C.}},
\oauthor{\bsnm{Wallraff}, \binits{A.}}:
{Low-Latency Digital Signal Processing for Feedback and Feedforward in Quantum Computing and Communication}.
Physical Review Applied
\textbf{9}(3)
(2018)
\doiurl{10.1103/PhysRevApplied.9.034011}
\end{botherref}
\endbibitem

\bibitem[\protect\citeauthoryear{Stefanazzi et~al.}{2022}]{Stefanazzi2022TheDetectors}
\begin{botherref}
\oauthor{\bsnm{Stefanazzi}, \binits{L.}},
\oauthor{\bsnm{Treptow}, \binits{K.}},
\oauthor{\bsnm{Wilcer}, \binits{N.}},
\oauthor{\bsnm{Stoughton}, \binits{C.}},
\oauthor{\bsnm{Bradford}, \binits{C.}},
\oauthor{\bsnm{Uemura}, \binits{S.}},
\oauthor{\bsnm{Zorzetti}, \binits{S.}},
\oauthor{\bsnm{Montella}, \binits{S.}},
\oauthor{\bsnm{Cancelo}, \binits{G.}},
\oauthor{\bsnm{Sussman}, \binits{S.}},
\oauthor{\bsnm{Houck}, \binits{A.}},
\oauthor{\bsnm{Saxena}, \binits{S.}},
\oauthor{\bsnm{Arnaldi}, \binits{H.}},
\oauthor{\bsnm{Agrawal}, \binits{A.}},
\oauthor{\bsnm{Zhang}, \binits{H.}},
\oauthor{\bsnm{Ding}, \binits{C.}},
\oauthor{\bsnm{Schuster}, \binits{D.I.}}:
{The QICK (Quantum Instrumentation Control Kit): Readout and control for qubits and detectors}.
Review of Scientific Instruments
\textbf{93}(4)
(2022)
\doiurl{10.1063/5.0076249}
\end{botherref}
\endbibitem

\bibitem[\protect\citeauthoryear{{Quantum Machines}}{2023}]{QuantumMachines2023OPX+:Hardware}
\begin{botherref}
\oauthor{\bsnm{{Quantum Machines}}}:
{OPX+: Ultra-Fast Quantum Control Hardware}
(2023).
\url{https://www.quantum-machines.co/products/opx/}
\end{botherref}
\endbibitem

\bibitem[\protect\citeauthoryear{{IBM}}{2022}]{IBM2022GoldeneyeSystem}
\begin{botherref}
\oauthor{\bsnm{{IBM}}}:
{Goldeneye cryogenic concept system}
(2022).
\url{https://research.ibm.com/blog/goldeneye-cryogenic-concept-system}
\end{botherref}
\endbibitem

\bibitem[\protect\citeauthoryear{Chen}{2023}]{Chen2023AreEfficient}
\begin{barticle}
\bauthor{\bsnm{Chen}, \binits{S.}}:
\batitle{{Are quantum computers really energy efficient?}}
\bjtitle{Nature Computational Science}
\bvolume{3}(\bissue{6}),
\bfpage{457}--\blpage{460}
(\byear{2023})
\doiurl{10.1038/s43588-023-00459-6}
\end{barticle}
\endbibitem

\bibitem[\protect\citeauthoryear{Fellous-Asiani et~al.}{2023}]{Fellous-Asiani2023OptimizingComputers}
\begin{barticle}
\bauthor{\bsnm{Fellous-Asiani}, \binits{M.}},
\bauthor{\bsnm{Chai}, \binits{J.H.}},
\bauthor{\bsnm{Thonnart}, \binits{Y.}},
\bauthor{\bsnm{Ng}, \binits{H.K.}},
\bauthor{\bsnm{Whitney}, \binits{R.S.}},
\bauthor{\bsnm{Auff{\`{e}}ves}, \binits{A.}}:
\batitle{{Optimizing Resource Efficiencies for Scalable Full-Stack Quantum Computers}}.
\bjtitle{PRX Quantum}
\bvolume{4}(\bissue{4}),
\bfpage{40319}
(\byear{2023})
\doiurl{10.1103/PRXQuantum.4.040319}
\end{barticle}
\endbibitem

\bibitem[\protect\citeauthoryear{Walter et~al.}{2017}]{Walter2017RapidQubits}
\begin{botherref}
\oauthor{\bsnm{Walter}, \binits{T.}},
\oauthor{\bsnm{Kurpiers}, \binits{P.}},
\oauthor{\bsnm{Gasparinetti}, \binits{S.}},
\oauthor{\bsnm{Magnard}, \binits{P.}},
\oauthor{\bsnm{Poto{\v{c}}nik}, \binits{A.}},
\oauthor{\bsnm{Salath{\'{e}}}, \binits{Y.}},
\oauthor{\bsnm{Pechal}, \binits{M.}},
\oauthor{\bsnm{Mondal}, \binits{M.}},
\oauthor{\bsnm{Oppliger}, \binits{M.}},
\oauthor{\bsnm{Eichler}, \binits{C.}},
\oauthor{\bsnm{Wallraff}, \binits{A.}}:
{Rapid High-Fidelity Single-Shot Dispersive Readout of Superconducting Qubits}.
Physical Review Applied
\textbf{7}(5)
(2017)
\doiurl{10.1103/PhysRevApplied.7.054020}
\end{botherref}
\endbibitem

\bibitem[\protect\citeauthoryear{DiVincenzo}{2009}]{DiVincenzo2009Fault-tolerantQubits}
\begin{bchapter}
\bauthor{\bsnm{DiVincenzo}, \binits{D.P.}}:
\bctitle{{Fault-tolerant architectures for superconducting qubits}}.
In: \bbtitle{Physica Scripta T},
vol. \bseriesno{T137}
(\byear{2009}).
\doiurl{10.1088/0031-8949/2009/T137/014020}
\end{bchapter}
\endbibitem

\bibitem[\protect\citeauthoryear{Barends et~al.}{2014}]{Barends2014SuperconductingTolerance}
\begin{barticle}
\bauthor{\bsnm{Barends}, \binits{R.}},
\bauthor{\bsnm{Kelly}, \binits{J.}},
\bauthor{\bsnm{Megrant}, \binits{A.}},
\bauthor{\bsnm{Veitia}, \binits{A.}},
\bauthor{\bsnm{Sank}, \binits{D.}},
\bauthor{\bsnm{Jeffrey}, \binits{E.}},
\bauthor{\bsnm{White}, \binits{T.C.}},
\bauthor{\bsnm{Mutus}, \binits{J.}},
\bauthor{\bsnm{Fowler}, \binits{A.G.}},
\bauthor{\bsnm{Campbell}, \binits{B.}},
\bauthor{\bsnm{Chen}, \binits{Y.}},
\bauthor{\bsnm{Chen}, \binits{Z.}},
\bauthor{\bsnm{Chiaro}, \binits{B.}},
\bauthor{\bsnm{Dunsworth}, \binits{A.}},
\bauthor{\bsnm{Neill}, \binits{C.}},
\bauthor{\bsnm{O'Malley}, \binits{P.}},
\bauthor{\bsnm{Roushan}, \binits{P.}},
\bauthor{\bsnm{Vainsencher}, \binits{A.}},
\bauthor{\bsnm{Wenner}, \binits{J.}},
\bauthor{\bsnm{Korotkov}, \binits{A.N.}},
\bauthor{\bsnm{Cleland}, \binits{A.N.}},
\bauthor{\bsnm{Martinis}, \binits{J.M.}}:
\batitle{{Superconducting quantum circuits at the surface code threshold for fault tolerance}}.
\bjtitle{Nature}
\bvolume{508}(\bissue{7497}),
\bfpage{500}--\blpage{503}
(\byear{2014})
\doiurl{10.1038/nature13171}
\end{barticle}
\endbibitem

\bibitem[\protect\citeauthoryear{Charbon et~al.}{2016}]{Charbon_CryoCMOSforQC}
\begin{bchapter}
\bauthor{\bsnm{Charbon}, \binits{E.}},
\bauthor{\bsnm{Sebastiano}, \binits{F.}},
\bauthor{\bsnm{Vladimirescu}, \binits{A.}},
\bauthor{\bsnm{Homulle}, \binits{H.}},
\bauthor{\bsnm{Visser}, \binits{S.}},
\bauthor{\bsnm{Song}, \binits{L.}},
\bauthor{\bsnm{Incandela}, \binits{R.M.}}:
\bctitle{Cryo-cmos for quantum computing}.
In: \bbtitle{2016 IEEE International Electron Devices Meeting (IEDM)},
pp. \bfpage{13}--\blpage{511354}
(\byear{2016}).
\doiurl{10.1109/IEDM.2016.7838410}
\end{bchapter}
\endbibitem

\bibitem[\protect\citeauthoryear{Sebastiano et~al.}{2020}]{Sebastiano2020Cryo-CMOSComputers}
\begin{bchapter}
\bauthor{\bsnm{Sebastiano}, \binits{F.}},
\bauthor{\bsnm{Van~Dijk}, \binits{J.P.G.}},
\bauthor{\bsnm{Thart}, \binits{P.A.}},
\bauthor{\bsnm{Patra}, \binits{B.}},
\bauthor{\bsnm{Van~Staveren}, \binits{J.}},
\bauthor{\bsnm{Xue}, \binits{X.}},
\bauthor{\bsnm{Almudever}, \binits{C.G.}},
\bauthor{\bsnm{Scappucci}, \binits{G.}},
\bauthor{\bsnm{Veldhorst}, \binits{M.}},
\bauthor{\bsnm{Vandersypen}, \binits{L.M.K.}},
\bauthor{\bsnm{Vladimirescu}, \binits{A.}},
\bauthor{\bsnm{Pellerano}, \binits{S.}},
\bauthor{\bsnm{Babaie}, \binits{M.}},
\bauthor{\bsnm{Charbon}, \binits{E.}}:
\bctitle{{Cryo-CMOS interfaces for large-scale quantum computers}}.
In: \bbtitle{Technical Digest - International Electron Devices Meeting, IEDM},
vol. \bseriesno{2020-December},
pp. \bfpage{1}--\blpage{25}.
\bpublisher{Institute of Electrical and Electronics Engineers Inc.}, \blocation{???}
(\byear{2020}).
\doiurl{10.1109/IEDM13553.2020.9372075}
\end{bchapter}
\endbibitem

\bibitem[\protect\citeauthoryear{Sebastiano et~al.}{2017}]{Sebastiano2017Cryo-CMOSInvited}
\begin{bchapter}
\bauthor{\bsnm{Sebastiano}, \binits{F.}},
\bauthor{\bsnm{Homulle}, \binits{H.}},
\bauthor{\bsnm{Patra}, \binits{B.}},
\bauthor{\bsnm{Incandela}, \binits{R.}},
\bauthor{\bsnm{Van~Dijk}, \binits{J.}},
\bauthor{\bsnm{Song}, \binits{L.}},
\bauthor{\bsnm{Babaie}, \binits{M.}},
\bauthor{\bsnm{Vladimirescu}, \binits{A.}},
\bauthor{\bsnm{Charbon}, \binits{E.}}:
\bctitle{{Cryo-CMOS Electronic Control for Scalable Quantum Computing: Invited}}.
In: \bbtitle{Proceedings - Design Automation Conference},
vol. \bseriesno{Part 128280}.
\bpublisher{Institute of Electrical and Electronics Engineers Inc.}, \blocation{???}
(\byear{2017}).
\doiurl{10.1145/3061639.3072948}
\end{bchapter}
\endbibitem

\bibitem[\protect\citeauthoryear{Patra et~al.}{2018}]{Patra2018Cryo-CMOSApplications}
\begin{barticle}
\bauthor{\bsnm{Patra}, \binits{B.}},
\bauthor{\bsnm{Incandela}, \binits{R.M.}},
\bauthor{\bsnm{Van~Dijk}, \binits{J.P.G.}},
\bauthor{\bsnm{Homulle}, \binits{H.A.R.}},
\bauthor{\bsnm{Song}, \binits{L.}},
\bauthor{\bsnm{Shahmohammadi}, \binits{M.}},
\bauthor{\bsnm{Staszewski}, \binits{R.B.}},
\bauthor{\bsnm{Vladimirescu}, \binits{A.}},
\bauthor{\bsnm{Babaie}, \binits{M.}},
\bauthor{\bsnm{Sebastiano}, \binits{F.}},
\bauthor{\bsnm{Charbon}, \binits{E.}}:
\batitle{{Cryo-CMOS Circuits and Systems for Quantum Computing Applications}}.
\bjtitle{IEEE Journal of Solid-State Circuits}
\bvolume{53}(\bissue{1}),
\bfpage{309}--\blpage{321}
(\byear{2018})
\doiurl{10.1109/JSSC.2017.2737549}
\end{barticle}
\endbibitem

\bibitem[\protect\citeauthoryear{Imroze et~al.}{2022}]{Imroze2022PackagedApplications}
\begin{bchapter}
\bauthor{\bsnm{Imroze}, \binits{F.}},
\bauthor{\bsnm{Nikbakhtnasrabadi}, \binits{F.}},
\bauthor{\bsnm{Danilin}, \binits{S.}},
\bauthor{\bsnm{Muhammad}, \binits{A.}},
\bauthor{\bsnm{Ahmad}, \binits{M.}},
\bauthor{\bsnm{Giagkoulovits}, \binits{C.}},
\bauthor{\bsnm{Heidari}, \binits{H.}},
\bauthor{\bsnm{Weides}, \binits{M.}}:
\bctitle{{Packaged CMOS cryogenic characterization for quantum computing applications}}.
In: \bbtitle{ICECS 2022 - 29th IEEE International Conference on Electronics, Circuits and Systems, Proceedings}.
\bpublisher{Institute of Electrical and Electronics Engineers Inc.}, \blocation{???}
(\byear{2022}).
\doiurl{10.1109/ICECS202256217.2022.9970996}
\end{bchapter}
\endbibitem

\bibitem[\protect\citeauthoryear{Giagkoulovits et~al.}{2022}]{Giagkoulovits2022CryoCMOSChallenges}
\begin{bchapter}
\bauthor{\bsnm{Giagkoulovits}, \binits{C.}},
\bauthor{\bsnm{Ahmad}, \binits{M.}},
\bauthor{\bsnm{Nikbakhtnasrabadi}, \binits{F.}},
\bauthor{\bsnm{Imroze}, \binits{F.}},
\bauthor{\bsnm{Weides}, \binits{M.}},
\bauthor{\bsnm{Heidari}, \binits{H.}}:
\bctitle{{CryoCMOS Characterization Strategies and Challenges}}.
In: \bbtitle{ICECS 2022 - 29th IEEE International Conference on Electronics, Circuits and Systems, Proceedings}.
\bpublisher{Institute of Electrical and Electronics Engineers Inc.}, \blocation{???}
(\byear{2022}).
\doiurl{10.1109/ICECS202256217.2022.9971010}
\end{bchapter}
\endbibitem

\bibitem[\protect\citeauthoryear{Bardin et~al.}{2019}]{8865458}
\begin{barticle}
\bauthor{\bsnm{Bardin}, \binits{J.C.}},
\bauthor{\bsnm{Jeffrey}, \binits{E.}},
\bauthor{\bsnm{Lucero}, \binits{E.}},
\bauthor{\bsnm{Huang}, \binits{T.}},
\bauthor{\bsnm{Das}, \binits{S.}},
\bauthor{\bsnm{Sank}, \binits{D.T.}},
\bauthor{\bsnm{Naaman}, \binits{O.}},
\bauthor{\bsnm{Megrant}, \binits{A.E.}},
\bauthor{\bsnm{Barends}, \binits{R.}},
\bauthor{\bsnm{White}, \binits{T.}},
\bauthor{\bsnm{Giustina}, \binits{M.}},
\bauthor{\bsnm{Satzinger}, \binits{K.J.}},
\bauthor{\bsnm{Arya}, \binits{K.}},
\bauthor{\bsnm{Roushan}, \binits{P.}},
\bauthor{\bsnm{Chiaro}, \binits{B.}},
\bauthor{\bsnm{Kelly}, \binits{J.}},
\bauthor{\bsnm{Chen}, \binits{Z.}},
\bauthor{\bsnm{Burkett}, \binits{B.}},
\bauthor{\bsnm{Chen}, \binits{Y.}},
\bauthor{\bsnm{Dunsworth}, \binits{A.}},
\bauthor{\bsnm{Fowler}, \binits{A.}},
\bauthor{\bsnm{Foxen}, \binits{B.}},
\bauthor{\bsnm{Gidney}, \binits{C.}},
\bauthor{\bsnm{Graff}, \binits{R.}},
\bauthor{\bsnm{Klimov}, \binits{P.}},
\bauthor{\bsnm{Mutus}, \binits{J.}},
\bauthor{\bsnm{McEwen}, \binits{M.J.}},
\bauthor{\bsnm{Neeley}, \binits{M.}},
\bauthor{\bsnm{Neill}, \binits{C.J.}},
\bauthor{\bsnm{Quintana}, \binits{C.}},
\bauthor{\bsnm{Vainsencher}, \binits{A.}},
\bauthor{\bsnm{Neven}, \binits{H.}},
\bauthor{\bsnm{Martinis}, \binits{J.}}:
\batitle{{Design and Characterization of a 28-nm Bulk-CMOS Cryogenic Quantum Controller Dissipating Less Than 2 mW at 3 K}}.
\bjtitle{{IEEE Journal of Solid-State Circuits}}
\bvolume{54}(\bissue{11}),
\bfpage{3043}--\blpage{3060}
(\byear{2019})
\doiurl{10.1109/JSSC.2019.2937234}
\end{barticle}
\endbibitem

\bibitem[\protect\citeauthoryear{Mehrpoo et~al.}{2019}]{Mehrpoo2019BenefitsComputers}
\begin{bchapter}
\bauthor{\bsnm{Mehrpoo}, \binits{M.}},
\bauthor{\bsnm{Patra}, \binits{B.}},
\bauthor{\bsnm{Gong}, \binits{J.}},
\bauthor{\bsnm{Dijk}, \binits{J.P.G.v.}},
\bauthor{\bsnm{Homulle}, \binits{H.}},
\bauthor{\bsnm{Kiene}, \binits{G.}},
\bauthor{\bsnm{Vladimirescu}, \binits{A.}},
\bauthor{\bsnm{Sebastiano}, \binits{F.}},
\bauthor{\bsnm{Charbon}, \binits{E.}},
\bauthor{\bsnm{Babaie}, \binits{M.}}:
\bctitle{{Benefits and Challenges of Designing Cryogenic CMOS RF Circuits for Quantum Computers}}.
In: \bbtitle{2019 IEEE International Symposium on Circuits and Systems (ISCAS)},
pp. \bfpage{1}--\blpage{5}
(\byear{2019}).
\doiurl{10.1109/ISCAS.2019.8702452}
\end{bchapter}
\endbibitem

\bibitem[\protect\citeauthoryear{Charbon et~al.}{2021}]{Charbon2021CryogenicProcessors}
\begin{barticle}
\bauthor{\bsnm{Charbon}, \binits{E.}},
\bauthor{\bsnm{Babaie}, \binits{M.}},
\bauthor{\bsnm{Vladimirescu}, \binits{A.}},
\bauthor{\bsnm{Sebastiano}, \binits{F.}}:
\batitle{{Cryogenic CMOS Circuits and Systems: Challenges and Opportunities in Designing the Electronic Interface for Quantum Processors}}.
\bjtitle{IEEE Microwave Magazine}
\bvolume{22}(\bissue{1}),
\bfpage{60}--\blpage{78}
(\byear{2021})
\doiurl{10.1109/MMM.2020.3023271}
\end{barticle}
\endbibitem

\bibitem[\protect\citeauthoryear{Pauka et~al.}{2021}]{Nayak2021FullTemperatures}
\begin{barticle}
\bauthor{\bsnm{Pauka}, \binits{S.J.}},
\bauthor{\bsnm{Das}, \binits{K.}},
\bauthor{\bsnm{Kalra}, \binits{R.}},
\bauthor{\bsnm{Moini}, \binits{A.}},
\bauthor{\bsnm{Yang}, \binits{Y.}},
\bauthor{\bsnm{Trainer}, \binits{M.}},
\bauthor{\bsnm{Bousquet}, \binits{A.}},
\bauthor{\bsnm{Cantaloube}, \binits{C.}},
\bauthor{\bsnm{Dick}, \binits{N.}},
\bauthor{\bsnm{Gardner}, \binits{G.C.}},
\bauthor{\bsnm{Manfra}, \binits{M.J.}},
\bauthor{\bsnm{Reilly}, \binits{D.J.}}:
\batitle{A cryogenic cmos chip for generating control signals for multiple qubits}.
\bjtitle{Nature Electronics}
\bvolume{4}(\bissue{1}),
\bfpage{64}--\blpage{70}
(\byear{2021})
\doiurl{10.1038/s41928-020-00528-y}
\end{barticle}
\endbibitem

\bibitem[\protect\citeauthoryear{Zhang et~al.}{2021}]{Zhang2021HotK}
\begin{barticle}
\bauthor{\bsnm{Zhang}, \binits{Y.}},
\bauthor{\bsnm{Xu}, \binits{J.}},
\bauthor{\bsnm{Lu}, \binits{T.T.}},
\bauthor{\bsnm{Zhang}, \binits{Y.}},
\bauthor{\bsnm{Luo}, \binits{C.}},
\bauthor{\bsnm{Guo}, \binits{G.}}:
\batitle{{Hot Carrier Degradation in MOSFETs at Cryogenic Temperatures down to 4.2 K}}.
\bjtitle{IEEE Transactions on Device and Materials Reliability}
\bvolume{21}(\bissue{4}),
\bfpage{620}--\blpage{626}
(\byear{2021})
\doiurl{10.1109/TDMR.2021.3124417}
\end{barticle}
\endbibitem

\bibitem[\protect\citeauthoryear{Beckers et~al.}{2018}]{Beckers2018CharacterizationK}
\begin{barticle}
\bauthor{\bsnm{Beckers}, \binits{A.}},
\bauthor{\bsnm{Jazaeri}, \binits{F.}},
\bauthor{\bsnm{Enz}, \binits{C.}}:
\batitle{{Characterization and Modeling of 28-nm Bulk CMOS Technology Down to 4.2 K}}.
\bjtitle{IEEE Journal of the Electron Devices Society}
\bvolume{6},
\bfpage{1007}--\blpage{1018}
(\byear{2018})
\doiurl{10.1109/JEDS.2018.2817458}
\end{barticle}
\endbibitem

\bibitem[\protect\citeauthoryear{Lu et~al.}{2020}]{Lu2020CharacterizationTemperature}
\begin{barticle}
\bauthor{\bsnm{Lu}, \binits{T.T.}},
\bauthor{\bsnm{Li}, \binits{Z.}},
\bauthor{\bsnm{Luo}, \binits{C.}},
\bauthor{\bsnm{Xu}, \binits{J.}},
\bauthor{\bsnm{Kong}, \binits{W.}},
\bauthor{\bsnm{Guo}, \binits{G.}}:
\batitle{{Characterization and Modeling of 0.18{$\mu$}m Bulk CMOS Technology at Sub-Kelvin Temperature}}.
\bjtitle{IEEE Journal of the Electron Devices Society}
\bvolume{8},
\bfpage{897}--\blpage{904}
(\byear{2020})
\doiurl{10.1109/JEDS.2020.3015265}
\end{barticle}
\endbibitem

\bibitem[\protect\citeauthoryear{Le~Guevel et~al.}{2020}]{LeGuevel2020Low-powerDevices}
\begin{botherref}
\oauthor{\bsnm{Le~Guevel}, \binits{L.}},
\oauthor{\bsnm{Billiot}, \binits{G.}},
\oauthor{\bsnm{Cardoso~Paz}, \binits{B.}},
\oauthor{\bsnm{Tagliaferri}, \binits{M.L.V.}},
\oauthor{\bsnm{De~Franceschi}, \binits{S.}},
\oauthor{\bsnm{Maurand}, \binits{R.}},
\oauthor{\bsnm{Cass{\'{e}}}, \binits{M.}},
\oauthor{\bsnm{Zurita}, \binits{M.}},
\oauthor{\bsnm{Sanquer}, \binits{M.}},
\oauthor{\bsnm{Vinet}, \binits{M.}},
\oauthor{\bsnm{Jehl}, \binits{X.}},
\oauthor{\bsnm{Jansen}, \binits{A.G.M.}},
\oauthor{\bsnm{Pillonnet}, \binits{G.}}:
{Low-power transimpedance amplifier for cryogenic integration with quantum devices}.
Applied Physics Reviews
\textbf{7}(4)
(2020)
\doiurl{10.1063/5.0007119}
\end{botherref}
\endbibitem

\bibitem[\protect\citeauthoryear{Yang et~al.}{2020}]{Yang2020AFDSOI}
\begin{barticle}
\bauthor{\bsnm{Yang}, \binits{Y.}},
\bauthor{\bsnm{Das}, \binits{K.}},
\bauthor{\bsnm{Moini}, \binits{A.}},
\bauthor{\bsnm{Reilly}, \binits{D.J.}}:
\batitle{{A Cryo-CMOS Voltage Reference in 28nm FDSOI}}.
\bjtitle{IEEE Solid-State Circuits Letters}
(\byear{2020})
\doiurl{10.1109/LSSC.2020.3010234}
\end{barticle}
\endbibitem

\bibitem[\protect\citeauthoryear{van Dijk et~al.}{2020}]{VanDijk_CryoCMOS_AnalogSignals}
\begin{bchapter}
\bauthor{\bsnm{Dijk}, \binits{J.}},
\bauthor{\bsnm{Hart}, \binits{P.}},
\bauthor{\bsnm{Kiene}, \binits{G.}},
\bauthor{\bsnm{Overwater}, \binits{R.}},
\bauthor{\bsnm{Padalia}, \binits{P.}},
\bauthor{\bsnm{Staveren}, \binits{J.}},
\bauthor{\bsnm{Babaie}, \binits{M.}},
\bauthor{\bsnm{Vladimirescu}, \binits{A.}},
\bauthor{\bsnm{Charbon}, \binits{E.}},
\bauthor{\bsnm{Sebastiano}, \binits{F.}}:
\bctitle{Cryo-cmos for analog/mixed-signal circuits and systems}.
In: \bbtitle{2020 IEEE Custom Integrated Circuits Conference (CICC)},
pp. \bfpage{1}--\blpage{8}
(\byear{2020}).
\doiurl{10.1109/CICC48029.2020.9075882}
\end{bchapter}
\endbibitem

\bibitem[\protect\citeauthoryear{Yoo et~al.}{2023}]{Yoo202334.2Unit-Cell}
\begin{bchapter}
\bauthor{\bsnm{Yoo}, \binits{J.}},
\bauthor{\bsnm{Chen}, \binits{Z.}},
\bauthor{\bsnm{Arute}, \binits{F.}},
\bauthor{\bsnm{Montazeri}, \binits{S.}},
\bauthor{\bsnm{Szalay}, \binits{M.}},
\bauthor{\bsnm{Erickson}, \binits{C.}},
\bauthor{\bsnm{Jeffrey}, \binits{E.}},
\bauthor{\bsnm{Fatemi}, \binits{R.}},
\bauthor{\bsnm{Giustina}, \binits{M.}},
\bauthor{\bsnm{Ansmann}, \binits{M.}},
\bauthor{\bsnm{Lucero}, \binits{E.}},
\bauthor{\bsnm{Kelly}, \binits{J.}},
\bauthor{\bsnm{Bardin}, \binits{J.C.}}:
\bctitle{{34.2 A 28-nm Bulk-CMOS IC for Full Control of a Superconducting Quantum Processor Unit-Cell}}.
In: \bbtitle{Digest of Technical Papers - IEEE International Solid-State Circuits Conference},
vol. \bseriesno{2023-February},
pp. \bfpage{506}--\blpage{508}.
\bpublisher{Institute of Electrical and Electronics Engineers Inc.}, \blocation{???}
(\byear{2023}).
\doiurl{10.1109/ISSCC42615.2023.10067292}
\end{bchapter}
\endbibitem

\bibitem[\protect\citeauthoryear{Park et~al.}{2021}]{Park2021AQubits}
\begin{barticle}
\bauthor{\bsnm{Park}, \binits{J.}},
\bauthor{\bsnm{Subramanian}, \binits{S.}},
\bauthor{\bsnm{Lampert}, \binits{L.}},
\bauthor{\bsnm{Mladenov}, \binits{T.}},
\bauthor{\bsnm{Klotchkov}, \binits{I.}},
\bauthor{\bsnm{Kurian}, \binits{D.J.}},
\bauthor{\bsnm{Juarez-Hernandez}, \binits{E.}},
\bauthor{\bsnm{Esparza}, \binits{B.P.}},
\bauthor{\bsnm{Kale}, \binits{S.R.}},
\bauthor{\bsnm{Asma~Beevi}, \binits{K.T.}},
\bauthor{\bsnm{Premaratne}, \binits{S.P.}},
\bauthor{\bsnm{Watson}, \binits{T.F.}},
\bauthor{\bsnm{Suzuki}, \binits{S.}},
\bauthor{\bsnm{Rahman}, \binits{M.}},
\bauthor{\bsnm{Timbadiya}, \binits{J.B.}},
\bauthor{\bsnm{Soni}, \binits{S.}},
\bauthor{\bsnm{Pellerano}, \binits{S.}}:
\batitle{{A Fully Integrated Cryo-CMOS SoC for State Manipulation, Readout, and High-Speed Gate Pulsing of Spin Qubits}}.
\bjtitle{IEEE Journal of Solid-State Circuits}
\bvolume{56}(\bissue{11}),
\bfpage{3289}--\blpage{3306}
(\byear{2021})
\doiurl{10.1109/JSSC.2021.3115988}
\end{barticle}
\endbibitem

\bibitem[\protect\citeauthoryear{Chakraborty et~al.}{2022}]{Chakraborty2022ATechnology}
\begin{barticle}
\bauthor{\bsnm{Chakraborty}, \binits{S.}},
\bauthor{\bsnm{Frank}, \binits{D.J.}},
\bauthor{\bsnm{Tien}, \binits{K.}},
\bauthor{\bsnm{Rosno}, \binits{P.}},
\bauthor{\bsnm{Yeck}, \binits{M.}},
\bauthor{\bsnm{Glick}, \binits{J.A.}},
\bauthor{\bsnm{Robertazzi}, \binits{R.}},
\bauthor{\bsnm{Richetta}, \binits{R.}},
\bauthor{\bsnm{Bulzacchelli}, \binits{J.F.}},
\bauthor{\bsnm{Underwood}, \binits{D.}},
\bauthor{\bsnm{Ramirez}, \binits{D.}},
\bauthor{\bsnm{Yilma}, \binits{D.}},
\bauthor{\bsnm{Davies}, \binits{A.}},
\bauthor{\bsnm{Joshi}, \binits{R.V.}},
\bauthor{\bsnm{Chambers}, \binits{S.D.}},
\bauthor{\bsnm{Lekuch}, \binits{S.}},
\bauthor{\bsnm{Inoue}, \binits{K.}},
\bauthor{\bsnm{Wisnieff}, \binits{D.}},
\bauthor{\bsnm{Baks}, \binits{C.W.}},
\bauthor{\bsnm{Bethune}, \binits{D.S.}},
\bauthor{\bsnm{Timmerwilke}, \binits{J.}},
\bauthor{\bsnm{Fox}, \binits{T.}},
\bauthor{\bsnm{Song}, \binits{P.}},
\bauthor{\bsnm{Johnson}, \binits{B.R.}},
\bauthor{\bsnm{Gaucher}, \binits{B.P.}},
\bauthor{\bsnm{Friedman}, \binits{D.J.}}:
\batitle{{A Cryo-CMOS Low-Power Semi-Autonomous Transmon Qubit State Controller in 14-nm FinFET Technology}}.
\bjtitle{IEEE Journal of Solid-State Circuits}
\bvolume{57}(\bissue{11}),
\bfpage{3258}--\blpage{3273}
(\byear{2022})
\doiurl{10.1109/JSSC.2022.3201775}
\end{barticle}
\endbibitem

\bibitem[\protect\citeauthoryear{{European Union}}{2024}]{ChipsJU_ARCTIC}
\begin{botherref}
\oauthor{\bsnm{{European Union}}}:
{European Chips Joint Undertaking: ARCTIC}.
\url{https://arctic-kdt.eu/}
(2024)
\end{botherref}
\endbibitem

\bibitem[\protect\citeauthoryear{Leonard et~al.}{2019}]{Leonard2019DigitalQubit}
\begin{botherref}
\oauthor{\bsnm{Leonard}, \binits{E.}},
\oauthor{\bsnm{Beck}, \binits{M.A.}},
\oauthor{\bsnm{Nelson}, \binits{J.}},
\oauthor{\bsnm{Christensen}, \binits{B.G.}},
\oauthor{\bsnm{Thorbeck}, \binits{T.}},
\oauthor{\bsnm{Howington}, \binits{C.}},
\oauthor{\bsnm{Opremcak}, \binits{A.}},
\oauthor{\bsnm{Pechenezhskiy}, \binits{I.V.}},
\oauthor{\bsnm{Dodge}, \binits{K.}},
\oauthor{\bsnm{Dupuis}, \binits{N.P.}},
\oauthor{\bsnm{Hutchings}, \binits{M.D.}},
\oauthor{\bsnm{Ku}, \binits{J.}},
\oauthor{\bsnm{Schlenker}, \binits{F.}},
\oauthor{\bsnm{Suttle}, \binits{J.}},
\oauthor{\bsnm{Wilen}, \binits{C.}},
\oauthor{\bsnm{Zhu}, \binits{S.}},
\oauthor{\bsnm{Vavilov}, \binits{M.G.}},
\oauthor{\bsnm{Plourde}, \binits{B.L.T.}},
\oauthor{\bsnm{McDermott}, \binits{R.}}:
{Digital Coherent Control of a Superconducting Qubit}.
Physical Review Applied
\textbf{11}(1)
(2019)
\doiurl{10.1103/PhysRevApplied.11.014009}
\end{botherref}
\endbibitem

\bibitem[\protect\citeauthoryear{Liu et~al.}{2023}]{Liu2023SingleModule}
\begin{botherref}
\oauthor{\bsnm{Liu}, \binits{C.H.}},
\oauthor{\bsnm{Ballard}, \binits{A.}},
\oauthor{\bsnm{Olaya}, \binits{D.}},
\oauthor{\bsnm{Schmidt}, \binits{D.R.}},
\oauthor{\bsnm{Biesecker}, \binits{J.}},
\oauthor{\bsnm{Lucas}, \binits{T.}},
\oauthor{\bsnm{Ullom}, \binits{J.}},
\oauthor{\bsnm{Patel}, \binits{S.}},
\oauthor{\bsnm{Rafferty}, \binits{O.}},
\oauthor{\bsnm{Opremcak}, \binits{A.}},
\oauthor{\bsnm{Dodge}, \binits{K.}},
\oauthor{\bsnm{Iaia}, \binits{V.}},
\oauthor{\bsnm{McBroom}, \binits{T.}},
\oauthor{\bsnm{Dubois}, \binits{J.L.}},
\oauthor{\bsnm{Hopkins}, \binits{P.F.}},
\oauthor{\bsnm{Benz}, \binits{S.P.}},
\oauthor{\bsnm{Plourde}, \binits{B.L.T.}},
\oauthor{\bsnm{McDermott}, \binits{R.}}:
{Single Flux Quantum-Based Digital Control of Superconducting Qubits in a Multichip Module}.
PRX Quantum
\textbf{4}(3)
(2023)
\doiurl{10.1103/PRXQuantum.4.030310}
\end{botherref}
\endbibitem

\bibitem[\protect\citeauthoryear{Likharev and Semenov}{1991}]{Likharev1991RSFQSystems}
\begin{botherref}
\oauthor{\bsnm{Likharev}, \binits{K.}},
\oauthor{\bsnm{Semenov}, \binits{V.}}:
{RSFQ Logic/ Memory Family: A New Josephson-Junction Technology for Sub-Terahertz-Clock-Frequency Digital Systems}.
IEEE Transactions on Applied Superconductivity
\textbf{1}(1)
(1991)
\end{botherref}
\endbibitem

\bibitem[\protect\citeauthoryear{Bunyk et~al.}{2001}]{Bunyk2001RSFQDevices}
\begin{barticle}
\bauthor{\bsnm{Bunyk}, \binits{P.}},
\bauthor{\bsnm{Likharev}, \binits{K.}},
\bauthor{\bsnm{Zinoviev}, \binits{D.}}:
\batitle{{RSFQ Technology: Physics and Devices}}.
\bjtitle{International Journal of High Speed Electronics and Systems}
\bvolume{11}(\bissue{01}),
\bfpage{257}--\blpage{305}
(\byear{2001})
\doiurl{10.1142/s012915640100085x}
\end{barticle}
\endbibitem

\bibitem[\protect\citeauthoryear{Chen et~al.}{1999}]{Chen1999RapidGHz}
\begin{botherref}
\oauthor{\bsnm{Chen}, \binits{W.}},
\oauthor{\bsnm{Rylyakov}, \binits{A.V.}},
\oauthor{\bsnm{Patel}, \binits{V.}},
\oauthor{\bsnm{Lukens}, \binits{J.E.}},
\oauthor{\bsnm{Likharev}, \binits{K.K.}}:
{Rapid Single Flux Quantum T-Flip Flop Operating up to 770 GHz}.
Technical Report~2
(1999)
\end{botherref}
\endbibitem

\bibitem[\protect\citeauthoryear{Barbosa et~al.}{2024}]{Barbosa2024}
\begin{botherref}
\oauthor{\bsnm{Barbosa}, \binits{J.}},
\oauthor{\bsnm{Brennan}, \binits{J.C.}},
\oauthor{\bsnm{Casaburi}, \binits{A.}},
\oauthor{\bsnm{Hutchings}, \binits{M.D.}},
\oauthor{\bsnm{Kirichenko}, \binits{A.}},
\oauthor{\bsnm{Mukhanov}, \binits{O.}},
\oauthor{\bsnm{Weides}, \binits{M.}}:
{RSFQ All-Digital Programmable Multi-Tone Generator For Quantum Applications}.
{IEEE Transactions on Quantum Engineering},
1--12
(2024)
\doiurl{10.1109/TQE.2024.3520805}
\end{botherref}
\endbibitem

\bibitem[\protect\citeauthoryear{McDermott and Vavilov}{2014}]{McDermott2014AccuratePulses}
\begin{botherref}
\oauthor{\bsnm{McDermott}, \binits{R.}},
\oauthor{\bsnm{Vavilov}, \binits{M.G.}}:
{Accurate Qubit Control with Single Flux Quantum Pulses}.
Physical Review Applied
\textbf{2}(1)
(2014)
\doiurl{10.1103/PhysRevApplied.2.014007}
\end{botherref}
\endbibitem

\bibitem[\protect\citeauthoryear{Geng et~al.}{2023}]{Geng2023QubitCircuits}
\begin{botherref}
\oauthor{\bsnm{Geng}, \binits{X.}},
\oauthor{\bsnm{Huang}, \binits{R.}},
\oauthor{\bsnm{He}, \binits{Y.}},
\oauthor{\bsnm{He}, \binits{K.}},
\oauthor{\bsnm{Dai}, \binits{G.}},
\oauthor{\bsnm{Yang}, \binits{L.}},
\oauthor{\bsnm{Wu}, \binits{X.}},
\oauthor{\bsnm{Yu}, \binits{Q.}},
\oauthor{\bsnm{Cheng}, \binits{M.}},
\oauthor{\bsnm{Chen}, \binits{G.}},
\oauthor{\bsnm{Liu}, \binits{J.}},
\oauthor{\bsnm{Chen}, \binits{W.}}:
{Qubit Energy Tuner Based on Single Flux Quantum Circuits}
(2023)
\doiurl{10.3389/fphy.2023.1215468}
\end{botherref}
\endbibitem

\bibitem[\protect\citeauthoryear{Li et~al.}{2019}]{Li2019Hardware-EfficientSequences}
\begin{botherref}
\oauthor{\bsnm{Li}, \binits{K.}},
\oauthor{\bsnm{McDermott}, \binits{R.}},
\oauthor{\bsnm{Vavilov}, \binits{M.G.}}:
{Hardware-Efficient Qubit Control with Single-Flux-Quantum Pulse Sequences}.
Physical Review Applied
\textbf{12}(1)
(2019)
\doiurl{10.1103/PhysRevApplied.12.014044}
\end{botherref}
\endbibitem

\bibitem[\protect\citeauthoryear{Jokar et~al.}{2021}]{Jokar2021PracticalGates}
\begin{bchapter}
\bauthor{\bsnm{Jokar}, \binits{M.R.}},
\bauthor{\bsnm{Rines}, \binits{R.}},
\bauthor{\bsnm{Chong}, \binits{F.T.}}:
\bctitle{{Practical implications of SFQ-based two-qubit gates}}.
In: \bbtitle{Proceedings - 2021 IEEE International Conference on Quantum Computing and Engineering, QCE 2021},
pp. \bfpage{402}--\blpage{412}.
\bpublisher{Institute of Electrical and Electronics Engineers Inc.}, \blocation{???}
(\byear{2021}).
\doiurl{10.1109/QCE52317.2021.00061}
\end{bchapter}
\endbibitem

\bibitem[\protect\citeauthoryear{Bernhardt et~al.}{2025}]{bernhardt2025quantumcomputercontrolledsuperconducting}
\begin{botherref}
\oauthor{\bsnm{Bernhardt}, \binits{J.}},
\oauthor{\bsnm{Jordan}, \binits{C.}},
\oauthor{\bsnm{Rahamim}, \binits{J.}},
\oauthor{\bsnm{Kirchenko}, \binits{A.}},
\oauthor{\bsnm{Bharadwaj}, \binits{K.}},
\oauthor{\bsnm{Fry-Bouriaux}, \binits{L.}},
\oauthor{\bsnm{Porsch}, \binits{K.}},
\oauthor{\bsnm{Somoroff}, \binits{A.}},
\oauthor{\bsnm{Tsai}, \binits{K.-T.}},
\oauthor{\bsnm{Walter}, \binits{J.}},
\oauthor{\bsnm{Weis}, \binits{A.}},
\oauthor{\bsnm{Yu}, \binits{M.-J.}},
\oauthor{\bsnm{Renzullo}, \binits{M.}},
\oauthor{\bsnm{Yohannes}, \binits{D.}},
\oauthor{\bsnm{Vernik}, \binits{I.}},
\oauthor{\bsnm{Mukhanov}, \binits{O.}},
\oauthor{\bsnm{Han}, \binits{S.J.}}:
Quantum Computer Controlled by Superconducting Digital Electronics at Millikelvin Temperature
(2025).
\url{https://arxiv.org/abs/2503.09879}
\end{botherref}
\endbibitem

\bibitem[\protect\citeauthoryear{Opremcak et~al.}{2018}]{Opremcak2018MeasurementCounter}
\begin{barticle}
\bauthor{\bsnm{Opremcak}, \binits{A.}},
\bauthor{\bsnm{Pechenezhskiy}, \binits{I.V.}},
\bauthor{\bsnm{Howington}, \binits{C.}},
\bauthor{\bsnm{Christensen}, \binits{B.G.}},
\bauthor{\bsnm{Beck}, \binits{M.A.}},
\bauthor{\bsnm{Leonard}, \binits{E.}},
\bauthor{\bsnm{Suttle}, \binits{J.}},
\bauthor{\bsnm{Wilen}, \binits{C.}},
\bauthor{\bsnm{Nesterov}, \binits{K.N.}},
\bauthor{\bsnm{Ribeill}, \binits{G.J.}},
\bauthor{\bsnm{Thorbeck}, \binits{T.}},
\bauthor{\bsnm{Schlenker}, \binits{F.}},
\bauthor{\bsnm{Vavilov}, \binits{M.G.}},
\bauthor{\bsnm{Plourde}, \binits{B.L.T.}},
\bauthor{\bsnm{McDermott}, \binits{R.}}:
\batitle{{Measurement of a superconducting qubit with a microwave photon counter}}.
\bjtitle{Science}
\bvolume{361}(\bissue{6408}),
\bfpage{1239}--\blpage{1242}
(\byear{2018})
\doiurl{10.1126/science.aat4625}
\end{barticle}
\endbibitem

\bibitem[\protect\citeauthoryear{Howington et~al.}{2019}]{Howington2019InterfacingReadout}
\begin{botherref}
\oauthor{\bsnm{Howington}, \binits{C.}},
\oauthor{\bsnm{Opremcak}, \binits{A.}},
\oauthor{\bsnm{McDermott}, \binits{R.}},
\oauthor{\bsnm{Kirichenko}, \binits{A.}},
\oauthor{\bsnm{Mukhanov}, \binits{O.A.}},
\oauthor{\bsnm{Plourde}, \binits{B.L.T.}}:
{Interfacing Superconducting Qubits with Cryogenic Logic: Readout}.
IEEE Transactions on Applied Superconductivity
\textbf{29}(5)
(2019)
\doiurl{10.1109/TASC.2019.2908884}
\end{botherref}
\endbibitem

\bibitem[\protect\citeauthoryear{Di~Palma et~al.}{2023}]{DiPalma2023DiscriminatingCircuit}
\begin{botherref}
\oauthor{\bsnm{Di~Palma}, \binits{L.}},
\oauthor{\bsnm{Miano}, \binits{A.}},
\oauthor{\bsnm{Mastrovito}, \binits{P.}},
\oauthor{\bsnm{Massarotti}, \binits{D.}},
\oauthor{\bsnm{Arzeo}, \binits{M.}},
\oauthor{\bsnm{Pepe}, \binits{G.P.}},
\oauthor{\bsnm{Tafuri}, \binits{F.}},
\oauthor{\bsnm{Mukhanov}, \binits{O.}}:
{Discriminating the Phase of a Coherent Tone with a Flux-Switchable Superconducting Circuit}.
Physical Review Applied
\textbf{19}(6)
(2023)
\doiurl{10.1103/PhysRevApplied.19.064025}
\end{botherref}
\endbibitem

\bibitem[\protect\citeauthoryear{Howe et~al.}{2022}]{Howe2022DigitalK}
\begin{botherref}
\oauthor{\bsnm{Howe}, \binits{L.}},
\oauthor{\bsnm{Castellanos-Beltran}, \binits{M.}},
\oauthor{\bsnm{Sirois}, \binits{A.J.}},
\oauthor{\bsnm{Olaya}, \binits{D.}},
\oauthor{\bsnm{Biesecker}, \binits{J.}},
\oauthor{\bsnm{Dresselhaus}, \binits{P.D.}},
\oauthor{\bsnm{Benz}, \binits{S.P.}},
\oauthor{\bsnm{Hopkins}, \binits{P.F.}}:
{Digital Control of a Superconducting Qubit Using a Josephson Pulse Generator at 3 K}.
PRX Quantum
\textbf{3}(1)
(2022)
\doiurl{10.1103/PRXQuantum.3.010350}
\end{botherref}
\endbibitem

\bibitem[\protect\citeauthoryear{McDermott et~al.}{2018}]{McDermott2018Quantum-classicalLogic}
\begin{botherref}
\oauthor{\bsnm{McDermott}, \binits{R.}},
\oauthor{\bsnm{Vavilov}, \binits{M.G.}},
\oauthor{\bsnm{Plourde}, \binits{B.L.T.}},
\oauthor{\bsnm{Wilhelm}, \binits{F.K.}},
\oauthor{\bsnm{Liebermann}, \binits{P.J.}},
\oauthor{\bsnm{Mukhanov}, \binits{O.A.}},
\oauthor{\bsnm{Ohki}, \binits{T.A.}}:
{Quantum-classical interface based on single flux quantum digital logic}.
Quantum Science and Technology
\textbf{3}(2)
(2018)
\doiurl{10.1088/2058-9565/aaa3a0}
\end{botherref}
\endbibitem

\bibitem[\protect\citeauthoryear{Takeuchi et~al.}{2024}]{aqfp24}
\begin{barticle}
\bauthor{\bsnm{Takeuchi}, \binits{N.}},
\bauthor{\bsnm{Yamae}, \binits{T.}},
\bauthor{\bsnm{Yamashita}, \binits{T.}},
\bauthor{\bsnm{Yamamoto}, \binits{T.}},
\bauthor{\bsnm{Yoshikawa}, \binits{N.}}:
\batitle{Microwave-multiplexed qubit controller using adiabatic superconductor logic}.
\bjtitle{npj Quantum Information}
\bvolume{10}(\bissue{1}),
\bfpage{53}
(\byear{2024})
\doiurl{10.1038/s41534-024-00849-2}
\end{barticle}
\endbibitem

\bibitem[\protect\citeauthoryear{Sinatkas et~al.}{2021}]{Sinatkas2021Electro-opticPhotonics}
\begin{botherref}
\oauthor{\bsnm{Sinatkas}, \binits{G.}},
\oauthor{\bsnm{Christopoulos}, \binits{T.}},
\oauthor{\bsnm{Tsilipakos}, \binits{O.}},
\oauthor{\bsnm{Kriezis}, \binits{E.E.}}:
{Electro-optic modulation in integrated photonics}.
American Institute of Physics Inc.
(2021).
\doiurl{10.1063/5.0048712}
\end{botherref}
\endbibitem

\bibitem[\protect\citeauthoryear{Weigel et~al.}{2018}]{Weigel2018BondedBandwidth}
\begin{barticle}
\bauthor{\bsnm{Weigel}, \binits{P.O.}},
\bauthor{\bsnm{Zhao}, \binits{J.}},
\bauthor{\bsnm{Fang}, \binits{K.}},
\bauthor{\bsnm{Al-Rubaye}, \binits{H.}},
\bauthor{\bsnm{Trotter}, \binits{D.}},
\bauthor{\bsnm{Hood}, \binits{D.}},
\bauthor{\bsnm{Mudrick}, \binits{J.}},
\bauthor{\bsnm{Dallo}, \binits{C.}},
\bauthor{\bsnm{Pomerene}, \binits{A.T.}},
\bauthor{\bsnm{Starbuck}, \binits{A.L.}},
\bauthor{\bsnm{DeRose}, \binits{C.T.}},
\bauthor{\bsnm{Lentine}, \binits{A.L.}},
\bauthor{\bsnm{Rebeiz}, \binits{G.}},
\bauthor{\bsnm{Mookherjea}, \binits{S.}}:
\batitle{{Bonded thin film lithium niobate modulator on a silicon photonics platform exceeding 100 GHz 3-dB electrical modulation bandwidth}}.
\bjtitle{Optics Express}
\bvolume{26}(\bissue{18}),
\bfpage{23728}
(\byear{2018})
\doiurl{10.1364/oe.26.023728}
\end{barticle}
\endbibitem

\bibitem[\protect\citeauthoryear{Youssefi et~al.}{2021}]{Youssefi2021ADevices}
\begin{barticle}
\bauthor{\bsnm{Youssefi}, \binits{A.}},
\bauthor{\bsnm{Shomroni}, \binits{I.}},
\bauthor{\bsnm{Joshi}, \binits{Y.J.}},
\bauthor{\bsnm{Bernier}, \binits{N.R.}},
\bauthor{\bsnm{Lukashchuk}, \binits{A.}},
\bauthor{\bsnm{Uhrich}, \binits{P.}},
\bauthor{\bsnm{Qiu}, \binits{L.}},
\bauthor{\bsnm{Kippenberg}, \binits{T.J.}}:
\batitle{{A cryogenic electro-optic interconnect for superconducting devices}}.
\bjtitle{Nature Electronics}
\bvolume{4}(\bissue{5}),
\bfpage{326}--\blpage{332}
(\byear{2021})
\doiurl{10.1038/s41928-021-00570-4}
\end{barticle}
\endbibitem

\bibitem[\protect\citeauthoryear{Lecocq et~al.}{2021}]{Lecocq2021ControlLink}
\begin{barticle}
\bauthor{\bsnm{Lecocq}, \binits{F.}},
\bauthor{\bsnm{Quinlan}, \binits{F.}},
\bauthor{\bsnm{Cicak}, \binits{K.}},
\bauthor{\bsnm{Aumentado}, \binits{J.}},
\bauthor{\bsnm{Diddams}, \binits{S.A.}},
\bauthor{\bsnm{Teufel}, \binits{J.D.}}:
\batitle{{Control and readout of a superconducting qubit using a photonic link}}.
\bjtitle{Nature}
\bvolume{591}(\bissue{7851}),
\bfpage{575}--\blpage{579}
(\byear{2021})
\doiurl{10.1038/s41586-021-03268-x}
\end{barticle}
\endbibitem

\bibitem[\protect\citeauthoryear{Yoshida et~al.}{1999}]{Yoshida1999AElectrodes}
\begin{botherref}
\oauthor{\bsnm{Yoshida}, \binits{K.}},
\oauthor{\bsnm{Kanda}, \binits{Y.}},
\oauthor{\bsnm{Kohjiro}, \binits{S.}}:
{A Traveling-Wave-Type LiNbO Optical Modulator with Superconducting Electrodes}.
Technical Report~7
(1999)
\end{botherref}
\endbibitem

\bibitem[\protect\citeauthoryear{de~Cea et~al.}{2020}]{deCea2020PhotonicDetectors}
\begin{botherref}
\oauthor{\bsnm{Cea}, \binits{M.}},
\oauthor{\bsnm{Wollman}, \binits{E.E.}},
\oauthor{\bsnm{Atabaki}, \binits{A.H.}},
\oauthor{\bsnm{Gray}, \binits{D.J.}},
\oauthor{\bsnm{Shaw}, \binits{M.D.}},
\oauthor{\bsnm{Ram}, \binits{R.J.}}:
{Photonic Readout of Superconducting Nanowire Single Photon Counting Detectors}.
Scientific Reports
\textbf{10}(1)
(2020)
\doiurl{10.1038/s41598-020-65971-5}
\end{botherref}
\endbibitem

\bibitem[\protect\citeauthoryear{Thiele et~al.}{2023}]{Thiele2023AllInterface}
\begin{botherref}
\oauthor{\bsnm{Thiele}, \binits{F.}},
\oauthor{\bsnm{Hummel}, \binits{T.}},
\oauthor{\bsnm{McCaughan}, \binits{A.N.}},
\oauthor{\bsnm{Brockmeier}, \binits{J.}},
\oauthor{\bsnm{Protte}, \binits{M.}},
\oauthor{\bsnm{Quiring}, \binits{V.}},
\oauthor{\bsnm{Lengeling}, \binits{S.}},
\oauthor{\bsnm{Eigner}, \binits{C.}},
\oauthor{\bsnm{Silberhorn}, \binits{C.}},
\oauthor{\bsnm{Bartley}, \binits{T.J.}}:
{All optical operation of a superconducting photonic interface}
(2023)
\end{botherref}
\endbibitem

\bibitem[\protect\citeauthoryear{Arnold et~al.}{2025}]{Arnold2025}
\begin{barticle}
\bauthor{\bsnm{Arnold}, \binits{G.}},
\bauthor{\bsnm{Werner}, \binits{T.}},
\bauthor{\bsnm{Sahu}, \binits{R.}},
\bauthor{\bsnm{Kapoor}, \binits{L.N.}},
\bauthor{\bsnm{Qiu}, \binits{L.}},
\bauthor{\bsnm{Fink}, \binits{J.M.}}:
\batitle{All-optical superconducting qubit readout}.
\bjtitle{Nature Physics}
\bvolume{21},
\bfpage{393}--\blpage{400}
(\byear{2025})
\doiurl{10.1038/s41567-024-02741-4}
\end{barticle}
\endbibitem

\bibitem[\protect\citeauthoryear{Al-Moathin et~al.}{2023}]{Al-Moathin2023CharacterizationApplications}
\begin{bchapter}
\bauthor{\bsnm{Al-Moathin}, \binits{A.}},
\bauthor{\bsnm{Zhong}, \binits{M.}},
\bauthor{\bsnm{Raghib Ali~Al-Taai}, \binits{Q.}},
\bauthor{\bsnm{Jiang}, \binits{Y.}},
\bauthor{\bsnm{Farage}, \binits{M.}},
\bauthor{\bsnm{Kazim}, \binits{J.U.R.}},
\bauthor{\bsnm{Ali}, \binits{M.}},
\bauthor{\bsnm{Nikbakhtnasrabadi}, \binits{F.}},
\bauthor{\bsnm{Powell}, \binits{M.}},
\bauthor{\bsnm{Khatri}, \binits{P.}},
\bauthor{\bsnm{Stanley}, \binits{M.}},
\bauthor{\bsnm{Rossi}, \binits{A.}},
\bauthor{\bsnm{Heidari}, \binits{H.}},
\bauthor{\bsnm{Imran}, \binits{M.}},
\bauthor{\bsnm{Abbasi}, \binits{Q.}},
\bauthor{\bsnm{Ridler}, \binits{N.}},
\bauthor{\bsnm{Weides}, \binits{M.}},
\bauthor{\bsnm{Li}, \binits{C.}}:
\bctitle{{Characterization of a Compact Wideband Microwave Metasurface Lens for Cryogenic Applications}}.
In: \bbtitle{2023 101st ARFTG Microwave Measurement Conference},
pp. \bfpage{1}--\blpage{4}
(\byear{2023}).
\doiurl{10.1109/ARFTG57476.2023.10279542}
\end{bchapter}
\endbibitem

\bibitem[\protect\citeauthoryear{Kazim et~al.}{2023}]{Kazim2023WirelessApplications}
\begin{bchapter}
\bauthor{\bsnm{Kazim}, \binits{J.U.R.}},
\bauthor{\bsnm{Ali}, \binits{M.}},
\bauthor{\bsnm{Al-Moathin}, \binits{A.}},
\bauthor{\bsnm{Nikbakhtnasrabadi}, \binits{F.}},
\bauthor{\bsnm{Khatri}, \binits{P.}},
\bauthor{\bsnm{Powell}, \binits{M.}},
\bauthor{\bsnm{Stanley}, \binits{M.}},
\bauthor{\bsnm{Ridler}, \binits{N.}},
\bauthor{\bsnm{Rossi}, \binits{A.}},
\bauthor{\bsnm{Weides}, \binits{M.}},
\bauthor{\bsnm{Heidari}, \binits{H.}},
\bauthor{\bsnm{Imran}, \binits{M.}},
\bauthor{\bsnm{Abbasi}, \binits{Q.}},
\bauthor{\bsnm{Li}, \binits{C.}}:
\bctitle{{Wireless Microwave Signal Transmission for Cryogenic Applications}}.
In: \bbtitle{2023 IEEE International Symposium on Antennas and Propagation and USNC-URSI Radio Science Meeting},
pp. \bfpage{43}--\blpage{44}
(\byear{2023}).
\doiurl{10.1109/USNC-URSI52151.2023.10237549}
\end{bchapter}
\endbibitem

\bibitem[\protect\citeauthoryear{Wang et~al.}{2023}]{Wang202334.1Interface}
\begin{bchapter}
\bauthor{\bsnm{Wang}, \binits{J.}},
\bauthor{\bsnm{Ibrahim}, \binits{M.I.}},
\bauthor{\bsnm{Harris}, \binits{I.B.}},
\bauthor{\bsnm{Monroe}, \binits{N.M.}},
\bauthor{\bsnm{Wasiq~Khan}, \binits{M.I.}},
\bauthor{\bsnm{Yi}, \binits{X.}},
\bauthor{\bsnm{Englund}, \binits{D.R.}},
\bauthor{\bsnm{Han}, \binits{R.}}:
\bctitle{{34.1 THz Cryo-CMOS Backscatter Transceiver: A Contactless 4 Kelvin-300 Kelvin Data Interface}}.
In: \bbtitle{Digest of Technical Papers - IEEE International Solid-State Circuits Conference},
vol. \bseriesno{2023-February},
pp. \bfpage{504}--\blpage{506}.
\bpublisher{Institute of Electrical and Electronics Engineers Inc.}, \blocation{???}
(\byear{2023}).
\doiurl{10.1109/ISSCC42615.2023.10067445}
\end{bchapter}
\endbibitem

\bibitem[\protect\citeauthoryear{Anders et~al.}{2023}]{Anders2023CMOSSciences}
\begin{barticle}
\bauthor{\bsnm{Anders}, \binits{J.}},
\bauthor{\bsnm{Babaie}, \binits{M.}},
\bauthor{\bsnm{Bardin}, \binits{J.}},
\bauthor{\bsnm{Bashir}, \binits{I.}},
\bauthor{\bsnm{Billiot}, \binits{G.}},
\bauthor{\bsnm{Blokhina}, \binits{E.}},
\bauthor{\bsnm{Bonen}, \binits{S.}},
\bauthor{\bsnm{Charbon}, \binits{E.}},
\bauthor{\bsnm{Chiaverini}, \binits{J.}},
\bauthor{\bsnm{Chuang}, \binits{I.}},
\bauthor{\bsnm{Degenhardt}, \binits{C.}},
\bauthor{\bsnm{Englund}, \binits{D.}},
\bauthor{\bsnm{Geck}, \binits{L.}},
\bauthor{\bsnm{Le~Guevel}, \binits{L.}},
\bauthor{\bsnm{Ham}, \binits{D.}},
\bauthor{\bsnm{Han}, \binits{R.}},
\bauthor{\bsnm{Ibrahim}, \binits{I.}},
\bauthor{\bsnm{Kr{\"{u}}ger}, \binits{D.}},
\bauthor{\bsnm{Lei}, \binits{K.M.}},
\bauthor{\bsnm{Voinigescu}, \binits{S.}}:
\batitle{{CMOS Integrated Circuits for the Quantum Information Sciences}}.
\bjtitle{IEEE Transactions on Quantum Engineering}
\bvolume{PP},
\bfpage{1}--\blpage{30}
(\byear{2023})
\doiurl{10.1109/TQE.2023.3290593}
\end{barticle}
\endbibitem

\bibitem[\protect\citeauthoryear{Alarcon et~al.}{2023}]{Alarcon2023ScalableNetwork-in-package}
\begin{bchapter}
\bauthor{\bsnm{Alarcon}, \binits{E.}},
\bauthor{\bsnm{Abadal}, \binits{S.}},
\bauthor{\bsnm{Sebastiano}, \binits{F.}},
\bauthor{\bsnm{Babaie}, \binits{M.}},
\bauthor{\bsnm{Charbon}, \binits{E.}},
\bauthor{\bsnm{Bolivar}, \binits{P.H.}},
\bauthor{\bsnm{Palesi}, \binits{M.}},
\bauthor{\bsnm{Blokhina}, \binits{E.}},
\bauthor{\bsnm{Leipold}, \binits{D.}},
\bauthor{\bsnm{Staszewski}, \binits{B.}},
\bauthor{\bsnm{Garcia-Saez}, \binits{A.}},
\bauthor{\bsnm{Almudever}, \binits{C.G.}}:
\bctitle{{Scalable multi-chip quantum architectures enabled by cryogenic hybrid wireless/quantum-coherent network-in-package}}.
In: \bbtitle{Proceedings - IEEE International Symposium on Circuits and Systems},
vol. \bseriesno{2023-May}.
\bpublisher{Institute of Electrical and Electronics Engineers Inc.}, \blocation{???}
(\byear{2023}).
\doiurl{10.1109/ISCAS46773.2023.10181857}
\end{bchapter}
\endbibitem

\bibitem[\protect\citeauthoryear{Anferov et~al.}{2024}]{anferov2024millimeterwavesuperconductingqubit}
\begin{botherref}
\oauthor{\bsnm{Anferov}, \binits{A.}},
\oauthor{\bsnm{Wan}, \binits{F.}},
\oauthor{\bsnm{Harvey}, \binits{S.P.}},
\oauthor{\bsnm{Simon}, \binits{J.}},
\oauthor{\bsnm{Schuster}, \binits{D.I.}}:
A Millimeter-Wave Superconducting Qubit
(2024).
\url{https://arxiv.org/abs/2411.11170}
\end{botherref}
\endbibitem

\bibitem[\protect\citeauthoryear{Wang et~al.}{2025}]{Wang2025}
\begin{barticle}
\bauthor{\bsnm{Wang}, \binits{J.}},
\bauthor{\bsnm{Harris}, \binits{I.}},
\bauthor{\bsnm{Ibrahim}, \binits{M.}},
\bauthor{\bsnm{Englund}, \binits{D.}},
\bauthor{\bsnm{Han}, \binits{R.}}:
\batitle{A wireless terahertz cryogenic interconnect that minimizes heat-to-information transfer}.
\bjtitle{Nature Electronics}
(\byear{2025})
\doiurl{10.1038/s41928-025-01355-9}
\end{barticle}
\endbibitem

\bibitem[\protect\citeauthoryear{Fellous-Asiani et~al.}{2023}]{PRXQuantum.4.040319}
\begin{barticle}
\bauthor{\bsnm{Fellous-Asiani}, \binits{M.}},
\bauthor{\bsnm{Chai}, \binits{J.H.}},
\bauthor{\bsnm{Thonnart}, \binits{Y.}},
\bauthor{\bsnm{Ng}, \binits{H.K.}},
\bauthor{\bsnm{Whitney}, \binits{R.S.}},
\bauthor{\bsnm{Auff\`eves}, \binits{A.}}:
\batitle{{Optimizing Resource Efficiencies for Scalable Full-Stack Quantum Computers}}.
\bjtitle{{PRX Quantum}}
\bvolume{4},
\bfpage{040319}
(\byear{2023})
\doiurl{10.1103/PRXQuantum.4.040319}
\end{barticle}
\endbibitem

\bibitem[\protect\citeauthoryear{Shen et~al.}{2024}]{Shen24photonicSFQ}
\begin{barticle}
\bauthor{\bsnm{Shen}, \binits{M.}},
\bauthor{\bsnm{Xie}, \binits{J.}},
\bauthor{\bsnm{Xu}, \binits{Y.}},
\bauthor{\bsnm{Wang}, \binits{S.}},
\bauthor{\bsnm{Cheng}, \binits{R.}},
\bauthor{\bsnm{Fu}, \binits{W.}},
\bauthor{\bsnm{Zhou}, \binits{Y.}},
\bauthor{\bsnm{Tang}, \binits{H.X.}}:
\batitle{Photonic link from single-flux-quantum circuits to room temperature}.
\bjtitle{Nature Photonics}
\bvolume{18}(\bissue{4}),
\bfpage{371}--\blpage{378}
(\byear{2024})
\doiurl{10.1038/s41566-023-01370-2}
\end{barticle}
\endbibitem

\bibitem[\protect\citeauthoryear{Jerger et~al.}{2012}]{freqmultiplex2012}
\begin{barticle}
\bauthor{\bsnm{Jerger}, \binits{M.}},
\bauthor{\bsnm{Poletto}, \binits{S.}},
\bauthor{\bsnm{Macha}, \binits{P.}},
\bauthor{\bsnm{Hübner}, \binits{U.}},
\bauthor{\bsnm{Il’ichev}, \binits{E.}},
\bauthor{\bsnm{Ustinov}, \binits{A.V.}}:
\batitle{Frequency division multiplexing readout and simultaneous manipulation of an array of flux qubits}.
\bjtitle{Applied Physics Letters}
\bvolume{101}(\bissue{4}),
\bfpage{042604}
(\byear{2012})
\doiurl{10.1063/1.4739454}
\end{barticle}
\endbibitem

\bibitem[\protect\citeauthoryear{Acharya et~al.}{2023}]{tdmultiplex2023}
\begin{barticle}
\bauthor{\bsnm{Acharya}, \binits{R.}},
\bauthor{\bsnm{Brebels}, \binits{S.}},
\bauthor{\bsnm{Grill}, \binits{A.}},
\bauthor{\bsnm{Verjauw}, \binits{J.}},
\bauthor{\bsnm{Ivanov}, \binits{T.}},
\bauthor{\bsnm{Lozano}, \binits{D.P.}},
\bauthor{\bsnm{Wan}, \binits{D.}},
\bauthor{\bsnm{Van~Damme}, \binits{J.}},
\bauthor{\bsnm{Vadiraj}, \binits{A.M.}},
\bauthor{\bsnm{Mongillo}, \binits{M.}},
\bauthor{\bsnm{Govoreanu}, \binits{B.}},
\bauthor{\bsnm{Craninckx}, \binits{J.}},
\bauthor{\bsnm{Radu}, \binits{I.P.}},
\bauthor{\bsnm{De~Greve}, \binits{K.}},
\bauthor{\bsnm{Gielen}, \binits{G.}},
\bauthor{\bsnm{Catthoor}, \binits{F.}},
\bauthor{\bsnm{Potočnik}, \binits{A.}}:
\batitle{Multiplexed superconducting qubit control at millikelvin temperatures with a low-power cryo-cmos multiplexer}.
\bjtitle{Nature Electronics}
\bvolume{6}(\bissue{11}),
\bfpage{900}--\blpage{909}
(\byear{2023})
\doiurl{10.1038/s41928-023-01033-8}
\end{barticle}
\endbibitem

\bibitem[\protect\citeauthoryear{{IBM Quantum}}{2024}]{IBMQuantumRoadmap}
\begin{botherref}
\oauthor{\bsnm{{IBM Quantum}}}:
The IBM Quantum Roadmap.
\url{https://www.ibm.com/quantum/blog/ibm-quantum-roadmap}.
Accessed: 2024-11-05
(2024)
\end{botherref}
\endbibitem

\end{thebibliography}

\end{document}